\documentclass[aps,pre,reprint,superscriptaddress,amsmath,amssymb,nofootinbib,floatfix]{revtex4-2}
\usepackage{graphicx,booktabs,bm}
\usepackage{microtype}
\usepackage[colorlinks=true,linkcolor=blue,citecolor=blue,urlcolor=blue]{hyperref}
\makeatletter
\AtBeginDocument{\let\auto@bib@innerbib\@empty
  \let\write@bibliographystyle\relax}
\makeatother
\hypersetup{pdftitle={Finite-time Mpemba crossings with imperfect preparation in reversible Markov systems},pdfauthor={Mesfin Asfaw Taye}}
\begin{document}
\title{Finite-time Mpemba crossings with imperfect preparation\texorpdfstring{\\}{ }
in reversible Markov systems}
\author{Mesfin Asfaw Taye}
\email{tayem@wlac.edu}
\affiliation{West Los Angeles College, Science Division,
9000 Overland Avenue, Culver City, California 90230, USA}
\date{October 24, 2026}

\begin{abstract}
Two preparations evolving under the same thermal Markov generator can
reverse their ordering in relative entropy even when the hotter state
retains a nonzero slow-mode component. We derive a sufficient finite-time
crossing condition for arbitrary finite irreducible reversible systems.
The condition compares the preparations' weights in the entire slowest
eigenspace, including degeneracies, and gives an explicit upper bound
using the separation from the next distinct decay rate. Exact
finite-amplitude entropy comparisons also yield all-mode enclosures and
bounds based on a retained spectrum with a controlled omitted tail.
A preparation-error inequality converts finite-time sign margins into
explicit probability tolerances; extensions account for temperature,
generator, and stationary-reference uncertainty. We test the bounds on
2000 predefined network draws with four graph families and 4--64 states.
There are 86 numerically located crossings, of which 45 satisfy the simple
sufficient condition. In a selected eight-state example, retaining two
nonstationary modes certifies a crossing before 1.55163, improving a
certified slow-sector endpoint of 2.53951; the located crossing is near
1.26943. A reflecting diffusion
in an asymmetric potential supplies a separate physical example, with
systematic mesh refinement. Independent interval arithmetic certifies
selected spectral bounds, finite-grid first crossings, and an entire preparation-temperature
rectangle in a three-state benchmark. The theorem does not require exact
slow-mode cancellation, while its numerical sharpness and detection
coverage remain model dependent. These results concern relative-entropy
relaxation and do not predict water-freezing times.
\end{abstract}
\maketitle

\section{Introduction}
\label{sec:introduction}

The Mpemba effect challenges the expectation that a preparation farther
from equilibrium must take longer to relax. Its original formulation
concerned hot and cold water \cite{mpemba1969,jeng2006}, but a controlled
statistical-mechanical question can be stated without phase change:
prepare two distributions and evolve them with the same final generator.
The hotter preparation may initially have a larger distance to the common
stationary state and later have a smaller one. Reversible Markov dynamics
provide a precise setting for this comparison \cite{luraz2017}.

The general spectral mechanism is already established. Relaxation rates
belong to the final generator, whereas the amplitudes of its modes depend
on the initial preparation. A reduced slow-mode overlap can produce
anomalous ordering, and cancellation of the slowest active sector gives
the strong Mpemba effect \cite{luraz2017,klich2019}. Anomalous relaxation
of this kind has been predicted in granular fluids
\cite{lasanta2017,biswas2020}, spin glasses \cite{baityjesi2019},
Langevin dynamics \cite{walker2021}, and open and many-body quantum systems
\cite{carollo2021,ares2023,moroder2024}, and observed in colloids
\cite{kumar2020,kumar2022} and trapped ions \cite{joshi2024,shapira2024}.
Inverse and precooling effects
\cite{gal2020,kumar2022,nava2025} and heating--cooling asymmetries
\cite{lapolla2020,vuhasegawa2021,meibohm2021,ibanez2024} belong to the
same broad family of relaxation speedups \cite{teza2026}.

Finite preparation accuracy raises a separate question. Exact removal of
a slow component is an ideal condition: a residual amplitude restores
that asymptotic decay rate. An ordinary finite-time crossing can
nevertheless persist. Establishing this persistence requires a comparison
over a specified time interval, rather than only an asymptotic exponent.
Preparation-error sensitivity has also been studied in quantum Mpemba
models \cite{mackinnon2026}, and finite-time criteria occur in quantum
imaginary-time evolution \cite{zeng2026}. The present setting is a
classical reversible semigroup with relative entropy as the specified
distance.

Relative entropy is attractive because, for a Gibbs stationary state, it
equals the excess nonequilibrium free energy divided by the bath
temperature. Its derivative is minus the dimensionless entropy-production
rate \cite{schnakenberg1976,seifert2012}, the central quantity of
stochastic thermodynamics \cite{Sekimoto2010,seifert2005}, whose exact
time dependence, together with the entropy-extraction rate, has been
obtained for Brownian particles, motors, and heat engines
\cite{Reimann2002,Hanggi2009,AsfawBekele2004,Asfaw2014ThermodynamicFeatureBrownian,Taye2016,Taye2020,Taye2021,Taye2022ExactTimeDependent2,Taye2024ExactTimeDependent,Taye2025ThermodynamicIrreversibilityUnderdamped,TayeExp2025}.
Entropy dissipation and functional inequalities already constrain
relaxation durations \cite{bobkov2006,diaconis1996,bao2025,auconi2026}. A global decay constant, however, does not
retain the different slow-mode weights of a particular pair of
preparations. Conversely, a quadratic entropy approximation is useful
near equilibrium but needs an explicit error bound before it can certify
a crossing at finite amplitude.

We use the established bounded-likelihood-ratio comparison between
relative entropy and chi-squared divergence \cite{sasonverdu2016}.
Combining this static inequality with the Markov maximum principle
provides exact time-dependent entropy envelopes. Their slow-sector form
gives a transparent sufficient condition, denoted $\Gamma>0$, and an
explicit time by which the initial ordering has reversed. Their all-mode
form retains more of the spectrum, while a nonnegative remainder bound
allows controlled use of only part of it. Degenerate slow eigenspaces are
treated as a whole.

We then quantify robustness. A centered entropy-gradient estimate and
probability contraction translate a prescribed preparation error into a
uniform entropy error. Positive separation throughout an early interval,
together with negative separation at a later endpoint, gives a robust
first-meeting interval. Temperature and rate uncertainty require
additional bounds, including the change of stationary reference when the
generator changes. The proof identifies these requirements explicitly.

Our main results are as follows. (i)~Theorem~1: for any finite reversible
chain, a positive slow-sector margin $\Gamma$ guarantees a crossing and
gives an explicit upper time, without exact slow-mode cancellation
(Fig.~\ref{fig:theorem}). (ii)~All-mode and partial-spectrum enclosures
that tighten this time using more spectral information, with a
controlled omitted tail. (iii)~Theorem~2 and its corollaries: explicit
tolerances for preparation, temperature, and rate errors, and a local
sensitivity formula expressed through the normalized entropy-production
rates. (iv)~Tests on $2000$ reversible networks with $4$ to $64$ states
and on a reflecting diffusion, together with interval-arithmetic
certificates, including a whole region of preparation temperatures
(Fig.~\ref{fig:benchmark_map}).

A companion manuscript develops the entropy-production-clock
interpretation and a tightly certified three-state benchmark, and states
the slow-sector comparison and preparation corollary in compact form
\cite{tayecompanion2026}. Here we give a self-contained treatment of the
general finite-state problem, the all-mode and partial-spectrum
construction, parameter robustness, predefined network ensembles, and a
physical diffusion model. The shared comparison ingredients are stated
explicitly; the additional scope and numerical tests are the focus of
this study.

The numerical work separates three kinds of evidence: evaluations in
ordinary floating-point arithmetic, mesh and algorithm comparisons, and
outward-rounded interval certificates. The network sample measures
performance within a declared ensemble, not the prevalence of the effect
among all Markov systems. Likewise, a finite-grid diffusion certificate
is not a proof of a continuum limit. The ordering considered throughout
is relative entropy; a simultaneous ordering of all monotones is a
different question addressed by thermomajorization \cite{vu2025}.

\section{Common dynamics, preparations, and thermodynamic distance}
\label{sec:setup}

\subsection{A fixed reversible Markov system}

Consider $N\ge2$ states with transition rates $k_{ij}$ for a jump from $i$ to
$j$. The probabilities are column vectors and obey
\begin{equation}
\dot p_i=\sum_{j\ne i}(k_{ji}p_j-k_{ij}p_i),\qquad \dot p=Wp.
\label{eq:master}
\end{equation}
Thus $W_{ij}=k_{ji}$ for $i\ne j$ and
$W_{ii}=-\sum_{j\ne i}k_{ij}$; every column of $W$ sums to zero.
We assume a finite, irreducible chain with a positive stationary
distribution $\pi$ and detailed balance,
\begin{equation}
W\pi=0,\qquad \pi_i k_{ij}=\pi_j k_{ji}.
\label{eq:db}
\end{equation}
The graph may be sparse, provided it is connected. Both preparations
evolve with precisely this same time-independent generator:
\begin{equation}
p^\alpha(t)=e^{Wt}p^\alpha(0),\qquad \alpha=H,C.
\label{eq:common_generator}
\end{equation}
The rates belong to the final dynamics; the preparations determine their
mode amplitudes. This is the standard spectral setting for the Markovian
Mpemba effect \cite{luraz2017,klich2019,vankampen2007,norris1997}.

The crossing bounds apply to arbitrary strictly positive initial
distributions satisfying the entropy ordering specified below. A thermal
realization has fixed energies $E_i$, bath temperature $T_b$, and
\begin{equation}
\pi_i=\frac{e^{-E_i/T_b}}{Z(T_b)},\qquad
p_i^\alpha(0)=\frac{e^{-E_i/T_\alpha}}{Z(T_\alpha)},
\quad T_H>T_C>T_b,
\label{eq:thermal_preparations}
\end{equation}
where $Z(T)=\sum_i e^{-E_i/T}$ and $k_{\rm B}=1$. Temperature labels refer
to these initial equilibria; an evolving nonequilibrium distribution
need not have a unique temperature.

\subsection{Entropy, a first meeting, and the normalized decay rate}

Use natural logarithms and the relative entropy
\begin{equation}
D_\alpha(t)=D[p^\alpha(t)\Vert\pi]
=\sum_i p_i^\alpha(t)\ln\frac{p_i^\alpha(t)}{\pi_i}.
\label{eq:Ddef}
\end{equation}
For a Gibbs stationary state, define
$\mathcal F[p]=\sum_i p_iE_i+T_b\sum_i p_i\ln p_i$. Then
\begin{equation}
\mathcal F[p]-\mathcal F[\pi]=T_bD[p\Vert\pi].
\label{eq:free_energy}
\end{equation}
Relative entropy therefore measures excess free energy in units of
$T_b$. For canonical cooling with nonconstant energies,
$D_H(0)>D_C(0)$ follows directly from $T_H>T_C>T_b$; the derivative
calculation is given in Appendix~\ref{app:foundations}.

With currents $J_{ij}=k_{ij}p_i-k_{ji}p_j$, the system entropy
$S=-\sum_ip_i\ln p_i$ obeys the balance
\begin{equation}
\begin{gathered}
\dot S=\dot e_p-\dot h_d,\qquad
\dot e_p=\frac12\sum_{i,j}J_{ij}\ln\frac{k_{ij}p_i}{k_{ji}p_j}\ge0,\\
\dot h_d=\frac12\sum_{i,j}J_{ij}\ln\frac{k_{ij}}{k_{ji}},
\end{gathered}
\label{eq:entropy_balance}
\end{equation}
where $\dot e_p$ is the dimensionless entropy-production rate and
$\dot h_d$ the entropy-extraction (heat-dissipation) rate into the bath
\cite{schnakenberg1976,seifert2005,seifert2012,Sekimoto2010}. Sums over
edges omit absent transitions. Under detailed balance
$\ln(k_{ij}/k_{ji})=(E_i-E_j)/T_b$, so $\dot h_d=-\dot U/T_b$, and since
$D=-S+U/T_b+\ln Z(T_b)$ the balance yields the standard identity
\begin{equation}
-\dot D_\alpha=\dot e_p^{\alpha},
\qquad r_\alpha=\frac{\dot e_p^{\alpha}}{D_\alpha}
=-\frac{\mathrm d}{\mathrm dt}\ln D_\alpha.
\label{eq:identity}
\end{equation}
Both initial entropies are assumed nonzero, and we impose
\begin{equation}
D_H(0)>D_C(0)>0,\qquad d(t)=D_H(t)-D_C(t).
\label{eq:initial_order}
\end{equation}
At any finite time a nonstationary preparation remains nonstationary,
because $e^{Wt}$ is invertible. Hence the rates in
Eq.~\eqref{eq:identity} remain defined.

Let
\begin{equation}
t_M=\inf\{t>0:d(t)=0\},
\label{eq:first_meeting}
\end{equation}
with $t_M=\infty$ when the set is empty; $t_M$ is the first-passage time
of the deterministic difference $d(t)$ through zero
\cite{Feller1971,CoxMiller1965,Duki2018StochasticResonanceFirst2,Taye2026StochasticFirstPassage}.
A meeting can be a tangency;
a Mpemba reversal additionally requires $d<0$ afterward. Our sufficient
theorem below guarantees a strict reversed ordering after a finite upper
bound, but does not assume uniqueness of earlier meetings.
Integrating Eq.~\eqref{eq:identity} gives
\begin{equation}
\begin{aligned}
\Theta_\alpha(t)&=\int_0^t r_\alpha(s)\,\mathrm ds
=\ln\frac{D_\alpha(0)}{D_\alpha(t)},\\
\ln\frac{D_H(t)}{D_C(t)}
&=\ln\frac{D_H(0)}{D_C(0)}
-\int_0^t(r_H-r_C)\,\mathrm ds.
\end{aligned}
\label{eq:clock}
\end{equation}
The dimensionless quantity $\Theta$ measures accumulated fractional
free-energy reduction. It is a progress variable constructed from the
trajectory, not an additional physical time coordinate. Connections
between entropy production and relaxation duration are well established
\cite{bao2025}; here the task is to bound the meeting of two specified
preparations. A relative-entropy crossing need not coincide with an
energy crossing or an ordering of every distance measure.

\section{Relaxation modes and exact finite-amplitude entropy envelopes}
\label{sec:spectral}

\subsection{The preparation weights}

Because $W\pi=0$, the deviation $p^\alpha(t)-\pi$ obeys the same master
equation and can be expanded in the eigenvectors of $W$,
\begin{equation}
p^\alpha(t)=\pi+\sum_{n=2}^Nb_{n,\alpha}\,e^{-\lambda_nt}\,v_n,
\qquad Wv_n=-\lambda_nv_n.
\label{eq:pexp}
\end{equation}
This is the whole mechanism in one line: the rates $\lambda_n$ belong to
the common generator, whereas the amplitudes $b_{n,\alpha}$ record how
strongly each preparation excites relaxation pattern $n$. Detailed
balance makes the patterns orthogonal once each component is weighted
by $1/\pi_i$, so we normalize $\sum_iv_{m,i}v_{n,i}/\pi_i=\delta_{mn}$.

For the entropy it is convenient to use the relative deviation
$f_i^\alpha=p_i^\alpha/\pi_i-1$, the fractional excess or deficit of
probability in state $i$, together with the equilibrium inner product
$\langle u,v\rangle_\pi=\sum_i\pi_i u_i v_i$. Normalization gives
$\langle f^\alpha,1\rangle_\pi=0$, and $f$ evolves with the
likelihood-ratio generator
\begin{equation}
(Lf)_i=\sum_{j\ne i}k_{ij}(f_j-f_i),\qquad \dot f=Lf,
\label{eq:Lgenerator}
\end{equation}
which detailed balance makes self-adjoint in this inner product. Its
eigenfunctions $\phi_{n,i}=v_{n,i}/\pi_i$ are normalized as
\begin{equation}
\begin{gathered}
L\phi_n=-\lambda_n\phi_n,\qquad
\langle\phi_m,\phi_n\rangle_\pi=\delta_{mn},\\
\phi_1=1,\qquad
0=\lambda_1<\lambda_2\le\cdots\le\lambda_N.
\end{gathered}
\label{eq:spectrum}
\end{equation}
The exact solution is
\begin{equation}
\begin{aligned}
f_i^\alpha(t)&=\sum_{n=2}^N b_{n,\alpha}e^{-\lambda_nt}\phi_{n,i},\\
b_{n,\alpha}&=\langle f^\alpha(0),\phi_n\rangle_\pi
=\sum_i p_i^\alpha(0)\phi_{n,i}.
\end{aligned}
\label{eq:fexp}
\end{equation}
The last equality uses the zero stationary mean of every nonstationary
eigenfunction. The weighted squared deviation is consequently
\begin{equation}
Q_\alpha(t)=\sum_i\pi_i[f_i^\alpha(t)]^2
=\sum_{n=2}^N b_{n,\alpha}^{\,2}e^{-2\lambda_nt}
=\chi^2(p^\alpha(t)\Vert\pi).
\label{eq:Qmodes}
\end{equation}
This is an exact spectral identity. Relative entropy is only
asymptotically $Q/2$ near equilibrium; replacing it by that approximation
at finite amplitude would leave an uncontrolled crossing-time error.
Appendix~\ref{app:foundations} derives the orthogonality and the expansion
\begin{equation}
D_\alpha(t)=\frac12Q_\alpha(t)
\left[1+O(\|f^\alpha(t)\|_\infty)\right].
\label{eq:quadratic_limit}
\end{equation}

\subsection{A scalar comparison that remains valid far from equilibrium}
\label{sec:envelopes}

Define
\begin{equation}
\begin{gathered}
g(x)=(1+x)\ln(1+x)-x,\qquad x>-1,\\
F(x)=\begin{cases}g(x)/x^2,&x\ne0,\\[1pt]1/2,&x=0.
\end{cases}
\end{gathered}
\label{eq:Fscalar}
\end{equation}
Since $g(0)=g'(0)=0$ and $g''(x)=1/(1+x)$, twice integrating gives
\begin{equation}
\begin{aligned}
F(x)&=\int_0^1\frac{1-u}{1+ux}\,\mathrm du>0,\\
F'(x)&=-\int_0^1\frac{u(1-u)}{(1+ux)^2}\,\mathrm du<0.
\end{aligned}
\label{eq:Fmonotone}
\end{equation}
Normalization cancels the linear term, so exactly
$D_\alpha=\sum_i\pi_i g(f_i^\alpha)
=\sum_i\pi_i(f_i^\alpha)^2F(f_i^\alpha)$.
For a reference time $s$, set
\begin{equation}
\begin{aligned}
m_\alpha(s)&=\min_i f_i^\alpha(s),
&M_\alpha(s)&=\max_i f_i^\alpha(s),\\
a_\alpha(s)&=F(M_\alpha(s)),
&c_\alpha(s)&=F(m_\alpha(s)).
\end{aligned}
\label{eq:range_factors}
\end{equation}
The likelihood-ratio dynamics maps every component at $t\ge s$ to a
convex combination of components at $s$:
\begin{equation}
f_i(t)=\sum_jK_{ij}(t-s)f_j(s),\qquad
K_{ij}(u)=[e^{Wu}]_{ij}\frac{\pi_j}{\pi_i},
\label{eq:ratio_kernel}
\end{equation}
where $K_{ij}\ge0$ and $\sum_jK_{ij}=1$. Therefore each later component
remains in $[m_\alpha(s),M_\alpha(s)]$. Monotonicity of $F$ gives
\begin{equation}
\boxed{a_\alpha(s)Q_\alpha(t)\le D_\alpha(t)
\le c_\alpha(s)Q_\alpha(t),\qquad t\ge s.}
\label{eq:envelopes}
\end{equation}
The static bounded-likelihood-ratio comparison between KL and
chi-squared divergences is established \cite{sasonverdu2016}. The
dynamical construction retains the actual preparation weights and uses
Markov contraction to keep the comparison valid over a time interval.
There is no omitted Taylor remainder. For fixed $s$ the factors remain
fixed; updating the reference time tightens them as the likelihood-ratio
range contracts toward zero.

\section{Finite-time crossing bounds with an occupied hot slow sector}
\label{sec:general_crossing}

\subsection{A sufficient condition for arbitrary finite state count}

Let $\lambda_s=\lambda_2$ and let $\lambda_f>\lambda_s$ be the next
\emph{distinct} positive rate. The explicit time formula in this
subsection assumes that this next rate exists. Define
\begin{equation}
\begin{gathered}
w_\alpha=\sum_{n:\lambda_n=\lambda_s}b_{n,\alpha}^{\,2},
\qquad q_H=\sum_{n:\lambda_n>\lambda_s}b_{n,H}^{\,2},\\
c_H=F(m_H(0)),\qquad a_C=F(M_C(0)),\\
\Gamma=a_Cw_C-c_Hw_H.
\end{gathered}
\label{eq:general_margin}
\end{equation}
The slow weight includes the entire eigenspace, so it is unchanged by
rotating the eigenbasis within a degeneracy. Physically, $w_H$ and $w_C$
measure the probability departure stored in the slowest sector; $q_H$
measures all faster hot contributions.

\emph{Theorem 1 (slow-sector crossing bound).}
For the common finite irreducible reversible generator and positive
preparations specified above, assume $D_H(0)>D_C(0)$ and $\Gamma>0$.
Then a first meeting exists and
\begin{equation}
\boxed{t_M\le T_{\rm up}=
\frac{1}{2(\lambda_f-\lambda_s)}
\ln\!\left(\frac{c_Hq_H}{\Gamma}\right).}
\label{eq:general_upper}
\end{equation}
Furthermore $D_H(t)<D_C(t)$ for every $t>T_{\rm up}$.

The proof is short enough to expose the physical content. Positivity of
the spectral weights gives
\begin{equation}
\begin{aligned}
Q_H(t)&\le w_He^{-2\lambda_st}+q_He^{-2\lambda_ft},\\
Q_C(t)&\ge w_Ce^{-2\lambda_st}.
\end{aligned}
\label{eq:sector_norm_bounds}
\end{equation}
Combining these with Eq.~\eqref{eq:envelopes} at $s=0$ yields
\begin{equation}
d(t)\le e^{-2\lambda_st}
\left[c_Hq_He^{-2(\lambda_f-\lambda_s)t}-\Gamma\right].
\label{eq:general_difference}
\end{equation}
Initially $d(0)>0$, so necessarily $c_Hq_H>\Gamma>0$ and
Eq.~\eqref{eq:general_upper} is positive and defined. Its time is exactly
when the bracket becomes zero, after which the bracket is negative.
Continuity proves the bound. Appendix~\ref{app:general_proofs} gives the
complete decomposition and explains restarting the bound at a later
reference time.

\begin{figure*}[t]
\centering
\includegraphics[width=\textwidth]{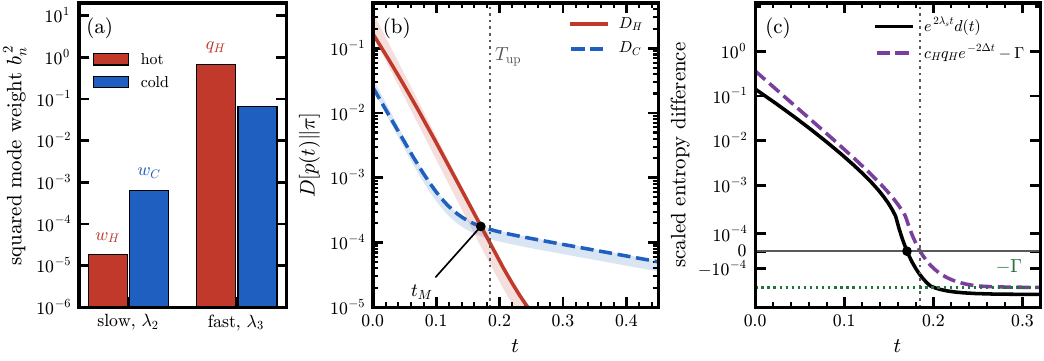}
\caption{How Theorem~1 works, for the three-state benchmark of
Sec.~\ref{sec:certified_robustness} at $(T_H,T_C)=(3,1.5)$, which is not an
exactly strong preparation. (a)~Squared mode weights: the hot preparation
has a much larger fast weight $q_H$ but a slow weight $w_H$ about $35$
times smaller than $w_C$. (b)~Relative entropies (lines) inside their exact
finite-amplitude envelopes, Eq.~\eqref{eq:envelopes} (shaded); the dot is
the crossing $t_M\simeq0.1702$ and the dotted line the bound
$T_{\rm up}\simeq0.1849$. (c)~The entropy difference multiplied by
$e^{2\lambda_st}$ (solid) and its upper bound from
Eq.~\eqref{eq:general_difference} (dashed). The bound becomes negative
exactly at $T_{\rm up}$ and tends to $-\Gamma$; the true difference
changes sign earlier. A symmetric-logarithmic vertical scale resolves
both signs.}
\label{fig:theorem}
\end{figure*}

The condition permits $w_H>0$; exact cancellation is unnecessary. The
initially larger hot distance can be dominated by fast contributions
that disappear before the colder preparation's slow contribution.
Figure~\ref{fig:theorem} shows this for a detuned three-state
preparation with $w_H\simeq1.81\times10^{-5}>0$,
$w_C\simeq6.41\times10^{-4}$, and $\Gamma\simeq2.17\times10^{-4}$: the
theorem guarantees a crossing by $T_{\rm up}\simeq0.185$, about $9\%$
after the actual crossing $t_M\simeq0.170$.
The positive margin $\Gamma$ quantifies the permitted hot slow weight
under this comparison, while $\lambda_f-\lambda_s$ controls how quickly
the excess fast contribution becomes small. Strong asymptotic
acceleration additionally requires $w_H=0$. For $w_H>0$ and $w_C>0$,
both preparations ultimately have the same exponential decay rate and
$D_H/D_C\to w_H/w_C$.

The theorem is sufficient and is not asserted to be uniformly sharp.
It discards faster cold contributions and merges all faster hot rates.
If $\Gamma\le0$, it decides nothing about crossing. If there is only one
distinct positive rate, this explicit formula does not apply; the
entropy envelopes still do. The theorem does not establish uniqueness of
meetings before $T_{\rm up}$. The familiar asymptotic condition
$w_H<w_C$, when $w_C>0$, is weaker than the displayed finite-amplitude
criterion; the latter supplies a finite upper time using the specified
range factors. The ingredients are the established spectral mechanism
\cite{luraz2017,klich2019} and divergence comparison
\cite{sasonverdu2016}; our claim concerns their preparation-dependent
finite-time use.

\subsection{All-mode bounds on the first meeting}

Keeping all weights gives tighter comparisons than merging the fast
sector. For any reference time $s$ define
\begin{equation}
\begin{aligned}
G_-(t;s)&=a_H(s)Q_H(t)-c_C(s)Q_C(t),\\
G_+(t;s)&=c_H(s)Q_H(t)-a_C(s)Q_C(t).
\end{aligned}
\end{equation}
The entropy envelopes imply
\begin{equation}
\boxed{G_-(t;s)\le d(t)\le G_+(t;s),\qquad t\ge s.}
\label{eq:general_envelopes}
\end{equation}
If earlier analysis excludes meetings before $s$, and $G_-(t;s)>0$
throughout $s\le t<L$, then $t_M\ge L$. Any $U\ge s$ with
$G_+(U;s)\le0$ gives $t_M\le U$. A lower bound on the \emph{first}
meeting requires positivity throughout the earlier interval; a positive
value at one endpoint does not exclude a previous meeting and recrossing.

For fixed $s$, these bounds are finite sums of exponentials.
Subdivision and interval evaluation can establish their signs. A
reference time can be advanced once the required early-time exclusion
has been proved. The reduced range factors then improve the entropy
comparison without changing the generator or initial amplitudes. Zeros
of the bounding exponential sums do not directly count the zeros of the
nonlinear entropy difference.

\subsection{A controlled remainder when only slow modes are known}
\label{sec:partial_spectrum}

Computing a full spectrum may be unnecessary. Retain modes $2,\ldots,m$
and suppose every omitted rate is at least
$\underline\lambda_{\rm cut}>0$. Define
\begin{equation}
\begin{aligned}
Q_\alpha^{\rm ret}(t)&=\sum_{n=2}^m b_{n,\alpha}^{\,2}e^{-2\lambda_nt},\\
R_\alpha(0)&=Q_\alpha(0)-\sum_{n=2}^m b_{n,\alpha}^{\,2}.
\end{aligned}
\end{equation}
All omitted squared weights are nonnegative, so
\begin{equation}
\boxed{0\le Q_\alpha(t)-Q_\alpha^{\rm ret}(t)
\le R_\alpha(0)e^{-2\underline\lambda_{\rm cut}t}.}
\label{eq:general_tail}
\end{equation}
The lower retained norm and upper retained-plus-remainder norm can be
substituted into Eq.~\eqref{eq:general_envelopes}, with a hot upper norm
and cold lower norm for an upper bound on $d$, and the reverse choices
for a lower bound. Dropping the omitted modes without this remainder
control does not provide a certificate.

Errors in the retained spectral data can be included explicitly. Suppose
certified intervals enclose
$\underline w_{n,\alpha}\le b_{n,\alpha}^2\le\overline w_{n,\alpha}$
and $\underline\lambda_n\le\lambda_n\le\overline\lambda_n$, with
nonnegative weight bounds and a certified
$R_\alpha(0)\le\overline R_\alpha$. Then for $t\ge0$,
\begin{equation}
\begin{aligned}
Q_\alpha^-(t)&=\sum_{n=2}^m
\underline w_{n,\alpha}e^{-2\overline\lambda_nt},\\
Q_\alpha^+(t)&=\sum_{n=2}^m
\overline w_{n,\alpha}e^{-2\underline\lambda_nt}
+\overline R_\alpha e^{-2\underline\lambda_{\rm cut}t},\\
Q_\alpha^-(t)&\le Q_\alpha(t)\le Q_\alpha^+(t).
\end{aligned}
\label{eq:partial_spectrum_error_bounds}
\end{equation}
For example, an upper bound on $Q_\alpha(0)$ minus the sum of retained
weight lower bounds gives a remainder upper bound. The range factors,
exponentials, and final sign comparisons must also be evaluated with
controlled rounding. Closely spaced or degenerate eigenvalues require
validated subspace weights or a justified orthonormal basis; a small
floating-point eigenvector residual alone is not an enclosure.

\section{Preparation tolerance and crossing-time sensitivity}
\label{sec:preparation_robustness}

\subsection{An explicit probability-error budget}

Keep the generator and stationary reference fixed. Let perturbed
preparations satisfy
\begin{equation}
\|\widetilde p^\alpha(0)-p^\alpha(0)\|_1\le\epsilon_\alpha.
\label{eq:preparation_error}
\end{equation}
Suppose all nominal and perturbed initial likelihood ratios lie in
$[h_{\alpha,-},h_{\alpha,+}]$, with $h_{\alpha,-}>0$. The stochastic
semigroup contracts the probability $1$-norm error, and its
likelihood-ratio kernel preserves the specified range. A centered
entropy-gradient estimate gives
\begin{equation}
\begin{gathered}
|\widetilde D_\alpha(t)-D_\alpha(t)|
\le C_\alpha\epsilon_\alpha,\qquad
C_\alpha=\frac12\ln\frac{h_{\alpha,+}}{h_{\alpha,-}},\\
|\widetilde d(t)-d(t)|\le
\delta_D:=C_H\epsilon_H+C_C\epsilon_C.
\end{gathered}
\label{eq:entropy_error}
\end{equation}
Appendix~\ref{app:robust_proofs} derives the factor $1/2$ by exploiting
normalization, rather than using an uncentered entropy gradient.

\emph{Theorem 2 (robust first-meeting interval).}
Suppose $0\le L<U$ and
\begin{equation}
d(t)\ge\eta_L>0\quad(0\le t\le L),\qquad
d(U)\le-\eta_U<0.
\label{eq:robust_sign_margins}
\end{equation}
Every allowed perturbed pair has its first meeting in $(L,U)$ if
\begin{equation}
\boxed{\delta_D<\min\{\eta_L,\eta_U\}.}
\label{eq:robust_error_budget}
\end{equation}
Indeed Eq.~\eqref{eq:entropy_error} preserves the positive early-time
gap and the negative endpoint gap. Continuity then places the first
zero strictly between them. The theorem establishes persistence with a
specified error budget; it does not require a perfectly canceled mode
or prove that the perturbed trajectory crosses only once.

The range assumption is checkable. Since a normalized probability error
with $1$-norm at most $\epsilon$ satisfies
$|\delta p_i|\le\epsilon/2$, one may use
\begin{equation}
\begin{aligned}
h_-&=\min_i\frac{p_i(0)-\epsilon/2}{\pi_i},\\
h_+&=\max_i\frac{p_i(0)+\epsilon/2}{\pi_i},
\end{aligned}
\label{eq:range_from_budget}
\end{equation}
provided $h_->0$. Direct bounds for the physical preparation family can
be less conservative. The same distinction applies to the sign margins:
they must be certified on the stated time set, rather than inferred
from a few sampled points.

\subsection{From temperature uncertainty to preparation uncertainty}

For the Gibbs preparations of Eq.~\eqref{eq:thermal_preparations},
\begin{equation}
\frac{\partial p_i(T)}{\partial T}
=\frac{p_i(T)[E_i-U(T)]}{T^2},\qquad
U(T)=\sum_i p_i(T)E_i.
\label{eq:Gibbs_derivative}
\end{equation}
The mean absolute energy deviation is at most half the energy range.
Consequently, for $T,\widetilde T\ge T_{\min}>0$,
\begin{equation}
\boxed{\|p(\widetilde T)-p(T)\|_1
\le\frac{E_{\max}-E_{\min}}{2T_{\min}^2}
|\widetilde T-T|.}
\label{eq:temperature_error}
\end{equation}
The proof uses Cauchy--Schwarz and an elementary bounded-variable
variance inequality, both shown in Appendix~\ref{app:robust_proofs}.
This converts a temperature interval into an explicit input for
Theorem 2. Interval evaluation of the Gibbs probabilities can improve
the numerical tolerance. A permitted temperature interval is a derived
conclusion only when its likelihood-ratio and sign-margin conditions
have also been verified.

\subsection{Transition-rate uncertainty changes the reference as well}
\label{sec:rate_robustness}

For two fixed irreducible generators $W$ and $\widetilde W$, put
$\delta_W=\|\widetilde W-W\|_{1\to1}$. Duhamel's formula and stochastic
contraction imply
\begin{equation}
\|\widetilde p^\alpha(t)-p^\alpha(t)\|_1
\le\epsilon_\alpha+t\delta_W.
\label{eq:generator_error}
\end{equation}
Let $C_\alpha$ now be a valid centered-gradient constant for both
distributions compared with the nominal $\pi$ over $0\le t\le U$.
If the perturbed stationary distribution is $\widetilde\pi$, define
$\delta_\pi=\max_i|\ln(\pi_i/\widetilde\pi_i)|$. Then
\begin{equation}
\begin{aligned}
|\widetilde d(t)-d(t)|
&\le\sum_{\alpha=H,C}C_\alpha(\epsilon_\alpha+U\delta_W)
+2\delta_\pi,\\[-2pt]
&\hspace{34mm}0\le t\le U.
\end{aligned}
\label{eq:rate_error_budget}
\end{equation}
Here $\widetilde D$ is measured relative to $\widetilde\pi$. If the
right-hand side is below both margins in
Eq.~\eqref{eq:robust_sign_margins}, the first meeting remains in $(L,U)$.
Appendix~\ref{app:rate_perturbations} supplies the proof and a way to
bound the likelihood-ratio ranges on the finite time interval. Keeping
the old stationary reference without the $\delta_\pi$ correction would
not establish the claimed uncertainty result.

\subsection{Local sensitivity and the entropy-production advantage}

For a smooth parameter $\theta$, consider a simple meeting root
$d(t_*(\theta),\theta)=0$. At the root write
$D_*=D_H(t_*)=D_C(t_*)>0$. The entropy balance gives
\begin{equation}
\partial_t d(t_*)=-D_*[r_H(t_*)-r_C(t_*)].
\end{equation}
When $r_H(t_*)\ne r_C(t_*)$, implicit differentiation therefore yields
\begin{equation}
\boxed{\frac{\partial t_*}{\partial\theta}
=\frac{\partial_\theta d(t_*,\theta)}
{D_*[r_H(t_*)-r_C(t_*)]}.}
\label{eq:crossing_sensitivity}
\end{equation}
The partial derivative holds time fixed and includes any dependence of
the stationary reference on $\theta$. For a fixed numerator, a small
normalized-rate difference makes the meeting time sensitive. This
tracks a local simple root; a first-meeting interpretation also needs
an early-time exclusion. At a tangent meeting the denominator vanishes,
and finite sign margins provide a separate route to robust conclusions.

\subsection{Scope and a negative control}

The spectral bound uses reversibility. The scalar entropy envelopes and
the fixed-generator probability-error argument need only a fixed
positive stationary distribution. For a nonreversible generator one
must evaluate the full semigroup norm for $Q$ instead of assuming
orthogonal eigenmodes. Chi-squared divergence remains contractive even
in that case; nonorthogonal modes do not imply transient growth of this
divergence. Appendix~\ref{app:scope} proves these statements and recalls
why nonadiabatic entropy production replaces total entropy production
in the fixed-reference clock \cite{esposito2010,hatano2001}.

A two-state Gibbs cooling protocol is a useful negative control. Its
excited-state probabilities remain ordered as
$p_1^H(t)>p_1^C(t)>\pi_1$ at every finite time, and binary relative
entropy increases with $p_1$ on that side of equilibrium. Hence it has
no relative-entropy crossing. The explicit proof in
Appendix~\ref{app:scope} applies to $T_H>T_C>T_b$, rather than to
arbitrary pairs of binary distributions.

\section{Tests on reversible network ensembles}
\label{sec:network-experiments}

\subsection{Sampling protocol and the quantities tested}

We test the finite-time condition in Eq.~\eqref{eq:general_margin}
without adjusting a preparation temperature to cancel a slow mode.
All network realizations use
\begin{equation}
(T_H,T_C,T_b)=(3,1.5,1),\qquad k_{\rm B}=1.
\label{eq:network-temperatures}
\end{equation}
For each $N\in\{4,8,16,32,64\}$ and each of four graph families, we
generate $100$ realizations. The $2000$ draws, including every unsuccessful
draw, are retained in the supplied data.

The energy levels are a random permutation of
\begin{equation}
E_j=\frac{5j}{N-1},\qquad j=0,\ldots,N-1.
\label{eq:network-energies}
\end{equation}
Thus every size has the same energy range; the placement of those levels
on the graph is randomized. Allowed undirected edges receive independent
symmetric prefactors
\begin{equation}
\gamma_{ij}=\gamma_{ji}=e^{X_{ij}},\qquad
X_{ij}\sim\operatorname{Uniform}[-4,2],
\label{eq:network-prefactors}
\end{equation}
and the directed rates are
\begin{equation}
k_{ij}=\gamma_{ij}e^{-(E_j-E_i)/2}.
\label{eq:network-rates}
\end{equation}
Detailed balance follows directly:
\begin{equation}
\pi_i k_{ij}
=\frac{\gamma_{ij}}{Z_b}e^{-(E_i+E_j)/2}
=\pi_j k_{ji}.
\end{equation}
The graph families are complete graphs, rings, birth--death paths, and
sparse connected graphs. A sparse graph contains a ring backbone, with
each remaining undirected edge added independently with probability
$\min(3/N,1)$. Each graph is therefore connected before any rate is drawn.
Appendix~\ref{sec:numerical-methods} specifies the random-number streams,
entropy evaluation, and crossing search.

For every draw we record the slow weights $w_H,w_C$, the sufficient
margin $\Gamma$, the upper formula $T_+=T_{\rm up}$ in
Eq.~\eqref{eq:general_upper},
and the outcome of a finite-time entropy search. Times are compared in
units $\tau=\lambda_s t$, which remove a uniform rescaling of all rates.
The search uses a declared time grid followed by root refinement. We call
its result a \emph{located crossing}; numerical root refinement alone
does not establish that this is the first or only crossing.

\subsection{Coverage and limitations of the sufficient condition}

Table~\ref{tab:network-counts} reports both successes and failures. There
are $86$ located crossings and $45$ positive-margin draws. The simple
condition therefore covers $45/86\simeq52.3\%$ of the crossings observed
in this sample. Every located crossing also has $w_H<w_C$. No transient
crossing with the opposite asymptotic ordering was detected by the
declared search. The sampling and finite search do not exclude that
possibility in another system.

\begin{table*}[t]
\centering
\caption{Located crossings / positive-$\Gamma$ draws, out of $100$
realizations in each cell, at the fixed temperatures in
Eq.~\eqref{eq:network-temperatures}. These are double-precision sample
counts. A negative margin is inconclusive, and failure to detect a
crossing in a finite search is not a no-crossing theorem.}
\label{tab:network-counts}
\begin{tabular}{rcccc}
\toprule
$N$ & Complete graph & Sparse connected & Ring & Birth--death\\
\midrule
4  & 0/0 & 0/0 & 0/0 & 0/0\\
8  & 1/1 & 3/1 & 1/1 & 4/2\\
16 & 4/2 & 6/2 & 4/2 & 12/5\\
32 & 6/2 & 5/3 & 5/3 & 10/7\\
64 & 3/2 & 7/4 & 3/2 & 12/6\\
\midrule
Total & 14/7 & 21/10 & 13/8 & 38/20\\
\bottomrule
\end{tabular}
\end{table*}

\begin{figure*}[t]
\centering
\includegraphics[width=\textwidth]{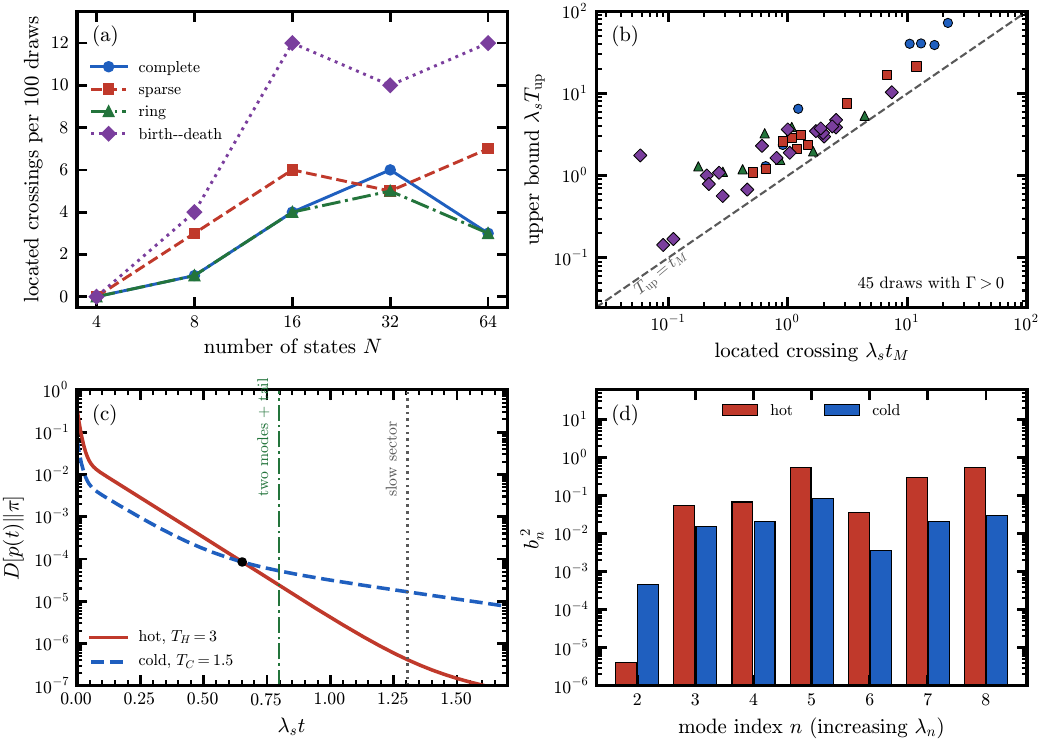}
\caption{Reversible network ensembles at fixed preparation temperatures
$(T_H,T_C,T_b)=(3,1.5,1)$. (a)~Located crossings per $100$ draws for four
graph families; lines guide the eye. (b)~Slow-sector upper bound
$T_{\rm up}$ against the located crossing $t_M$ for the $45$ draws with
$\Gamma>0$, in units of $\lambda_s$; the dashed diagonal marks equality,
so every point lies on the conservative side. (c)~Relative entropies of
the selected eight-state complete graph. The dot marks the crossing; the
dotted and dash-dotted lines are the slow-sector bound and the bound
retaining two modes plus a controlled tail. (d)~Squared weights of all
seven nonstationary modes: the hot slow weight $b_2^2$ is small but
nonzero. All curves are double-precision evaluations; the certified
statements are listed in Table~\ref{tab:certified-models}.}
\label{fig:network-study}
\end{figure*}

Among the $45$ positive-margin draws, the ratio of the upper formula to
the located crossing ranges from $1.214$ to $30.193$, with median $2.154$.
The criterion is useful as a sufficient guarantee, but it can be loose.
In particular, no dimension-independent sharpness claim follows from
Fig.~\ref{fig:network-study}. The occurrence counts also depend on the
chosen energy range, temperature pair, graph ensemble, and rate
distribution. The absence of an observed crossing in the sampled
four-state graphs does not imply a four-state impossibility result.

\subsection{An eight-state example and a partial-spectrum improvement}

To fix a selection rule independently of the subsequent plots, we use
the first positive-margin draw in the family--size--draw order specified
in Appendix~\ref{sec:numerical-methods}. It is complete graph $N=8$,
draw $6$. The energy vector is
\begin{equation}
E=\frac57(5,2,3,0,7,6,1,4).
\label{eq:selected-network-energy}
\end{equation}
The complete symmetric prefactor matrix is supplied as exact finite
decimals in \texttt{selected\_network\_exact\_inputs.json}; this file,
Eq.~\eqref{eq:network-rates}, and the energy vector define the reference
model unambiguously. Its first rates and weights are
\begin{align}
\lambda_s&\simeq0.51374978,&\lambda_f&\simeq2.27730802,\\
w_H&\simeq4.1466755\times10^{-6},&w_C&\simeq4.5240625\times10^{-4},\\
q_H&\simeq1.5733603,&\Gamma&\simeq1.2529349\times10^{-4}.
\end{align}
The initial entropies are $D_H(0)\simeq0.38363557$ and
$D_C(0)\simeq0.06814960$. The located crossing and evaluated upper formula
are
\begin{equation}
t_{\rm cross}\simeq1.2694268,\qquad T_+\simeq2.5395000.
\label{eq:selected-network-times}
\end{equation}
An independent interval matrix exponential, together with a semigroup
enclosure proving positivity throughout $[0,1.26]$, certifies the
\emph{first} crossing in $(1.26,1.28)$ for the exact reference inputs.
Appendix~\ref{app:certification} supplies the interval-cover proof.
This propagation proof is independent of a separate validated spectral
calculation. The latter encloses the simple spectrum and its eigenvectors,
proves $w_H>0$ and $\Gamma>0$, and establishes the outward upper endpoint
\begin{equation}
T_{\rm up}<2.53950003286112.
\end{equation}
Neither calculation asserts global uniqueness of crossings.

The partial-spectrum bound of Eq.~\eqref{eq:general_tail} gives a useful
improvement. Keep the first two nonstationary modes and denote their
weights by $b_{2,\alpha}^2,b_{3,\alpha}^2$. The next omitted rate is
$\lambda_4\simeq17.27305636$. The hot norm is bounded above by
\begin{align}
Q_H(t)\le{}&b_{2,H}^2e^{-2\lambda_2t}
+b_{3,H}^2e^{-2\lambda_3t}\nonumber\\
&+R_H(0)e^{-2\lambda_4t},
\label{eq:selected-partial-hot}
\end{align}
where
$R_H(0)=Q_H(0)-b_{2,H}^2-b_{3,H}^2$. Dropping the nonnegative cold
remainder supplies the corresponding lower bound for $Q_C(t)$. Inserting
these into $c_HQ_H-a_CQ_C$ gives an evaluated crossing upper bound of
$1.5516211$. The full-spectrum envelope gives the same value at the
displayed precision. Thus a bound on the omitted contribution can retain
most of the useful information in this example. Here the full spectrum
was computed to evaluate and check the partial-information formula; we
do not claim a measured computational speedup. A separate interval
calculation validates both retained weights, their rates, and the
omitted tail, using the exact lower cutoff
$\underline\lambda_{\rm cut}=17.27305635$. At the exact time
$t=1.55163$, the resulting upper envelope is strictly negative,
\begin{equation}
G_+(1.55163;0)<-8.000112\times10^{-10}<0.
\end{equation}
Thus $t_M<1.55163$ is a rigorous partial-spectrum upper bound for the
specified eight-state model. The decimal $1.5516211$ above is a numerical
root estimate for the envelope, not the endpoint used in this certificate.

\section{A physical example: reflecting diffusion in an asymmetric potential}
\label{sec:diffusion-experiments}

\subsection{Dynamics, parameter search, and selection}

Consider overdamped diffusion on the bounded interval $[-3,3]$,
\begin{equation}
\partial_t\rho(x,t)=\partial_x[U'(x)\rho(x,t)+\partial_x\rho(x,t)],
\label{eq:diffusion-FPE}
\end{equation}
with reflecting boundaries and units $T_b=\gamma=k_{\rm B}=1$.
Both preparations evolve with this same final potential and bath. Their
initial densities are proportional to $e^{-U(x)/T_H}$ and
$e^{-U(x)/T_C}$, respectively. We use
\begin{equation}
U(x)=\frac{a x^2}{2}
-A e^{-(x+1)^2/(2s_L^2)}
-B e^{-(x-1)^2/(2s_R^2)}.
\label{eq:diffusion-potential}
\end{equation}

\begin{figure*}[t]
\centering
\includegraphics[width=\textwidth]{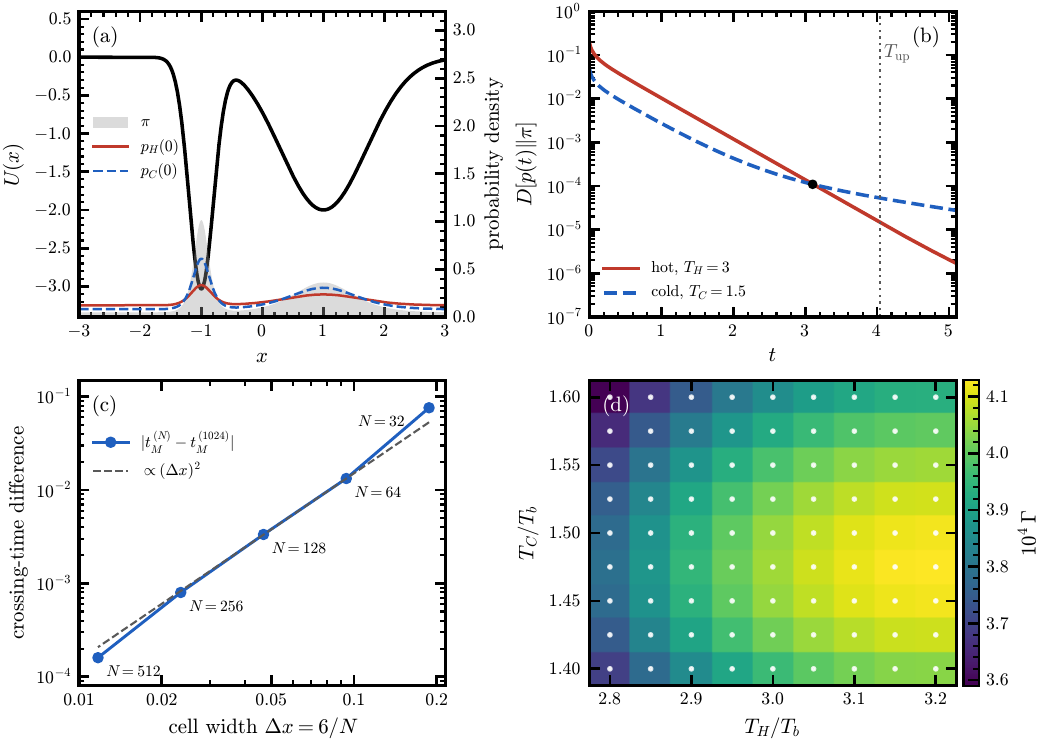}
\caption{Reflecting diffusion in the selected asymmetric potential.
(a)~Potential $U(x)$ on $[-3,3]$ (left axis) with the stationary density
and the two initial Gibbs densities (right axis). (b)~Relative entropies
at $N=512$; the dot marks the crossing and the dotted line the
slow-sector bound $T_{\rm up}$. (c)~Difference between the crossing on each
grid and on the finest grid, with a $(\Delta x)^2$ guide. (d)~Margin
$\Gamma$ at all $81$ sampled temperature pairs (white points); the
colored cells only visualize these point values.}
\label{fig:diffusion-study}
\end{figure*}

The finite search takes $A\in\{3,4,5,6\}$,
$B\in\{2,3,4,5\}$ with $A>B$, $s_L\in\{0.2,0.3,0.5\}$,
$s_R\in\{0.4,0.7,1\}$, and $a\in\{0,0.5\}$, in that
Cartesian-product order. At $N=128$ this gives $180$ potential choices.
Seven show a located crossing and have $w_H<w_C$; three satisfy
$\Gamma>0$. We select the first positive-margin case:
\begin{equation}
(A,B,s_L,s_R,a)=(3,2,0.2,0.7,0).
\label{eq:diffusion-selected}
\end{equation}
The temperatures remain $(T_H,T_C,T_b)=(3,1.5,1)$ throughout. The selected
potential has a narrow deeper left well and a broader right well
[Fig.~\ref{fig:diffusion-study}(a)]. Selection from a disclosed finite
parameter search is different from a prevalence statement; neither
temperature is found by solving a slow-mode zero.

\subsection{A reversible discretization and mesh refinement}

Let $x_i=-3+(i+1/2)\Delta x$, $i=0,\ldots,N-1$, with $\Delta x=6/N$.
Only nearest neighbors communicate, with
\begin{align}
k_{i,i+1}&=(\Delta x)^{-2}e^{-[U(x_{i+1})-U(x_i)]/2},\nonumber\\
k_{i+1,i}&=(\Delta x)^{-2}e^{+[U(x_{i+1})-U(x_i)]/2}.
\label{eq:diffusion-rates}
\end{align}
No jumps leave the interval. The finite model obeys detailed balance
with sampled Gibbs weights. Appendix~\ref{sec:numerical-methods} derives
the flux consistency with Eq.~\eqref{eq:diffusion-FPE}. The finite-state
theorems apply to every such discretization.

Table~\ref{tab:diffusion-mesh} shows that the located crossing and upper
formula stabilize under refinement. The difference between the $N=512$
and $N=1024$ located crossings is $1.6005\times10^{-4}$, approximately
$0.0052\%$ of the finer value. This is a numerical convergence check;
it is not an interval bound on the continuum crossing time. In
particular, no extrapolated continuum value is used as a certified result.

At $N=512$, the computed slow weights are
$w_H\simeq1.1371813\times10^{-5}$ and
$w_C\simeq9.8849729\times10^{-4}$. The hot contribution is small and
nonzero at numerical precision. For the exact $64$-cell discretization,
independent interval propagation gives
\begin{equation}
D_H(3)-D_C(3)>0,\qquad
D_H(3.2)-D_C(3.2)<0.
\label{eq:diffusion-certified-signs}
\end{equation}
An additional semigroup enclosure proves $D_H(t)-D_C(t)>0$ throughout
$0\le t\le3$. Together these results certify the \emph{first} crossing
in $(3,3.2)$ for that finite model. The interval-cover proof is given in
Appendix~\ref{app:certification}. A separate validated eigensystem
also establishes $w_H>0$, $\Gamma>0$, and
$T_{\rm up}<4.02143847094728$ for this grid. Global uniqueness and the
continuum limit are not certified.

\begin{table}[htbp]
\centering
\caption{Mesh-refinement evidence for the selected reflecting diffusion.
Each row is a double-precision evaluation for its finite grid. The
certificate in Appendix~\ref{app:certification} applies separately to
$N=64$.}
\label{tab:diffusion-mesh}
\begin{tabular}{rccc}
\toprule
$N$ & $\lambda_s$ & Located crossing & $T_+$\\
\midrule
32   & 0.29133854 & 3.0278394 & 3.9513325\\
64   & 0.28608283 & 3.0905462 & 4.0214385\\
128  & 0.28461341 & 3.1004873 & 4.0337451\\
256  & 0.28424130 & 3.1030296 & 4.0371088\\
512  & 0.28414797 & 3.1036689 & 4.0378437\\
1024 & 0.28412462 & 3.1038290 & 4.0380828\\
\bottomrule
\end{tabular}
\end{table}

\subsection{Preparation variation and a negative control}

A separate $9\times9$ pointwise survey at $N=128$ takes
$T_H\in[2.8,3.2]$ and $T_C\in[1.4,1.6]$, with equally spaced values on
each axis. Every one of the $81$ pairs has a positive computed margin,
between $3.5897\times10^{-4}$ and $4.1289\times10^{-4}$, and a located
crossing between $3.01944$ and $3.21891$. This supports robustness of the
observed diffusion example over the sampled preparations. It does not
certify the full continuous temperature rectangle. The separate
three-state rectangle in Appendix~\ref{app:certification} does carry
that stronger interval guarantee.

For an analytical negative control, take a two-state Gibbs system with
energies $(0,2)$, symmetric prefactor one, and the same temperatures.
Its excited-state occupation satisfies
\begin{equation}
p_1^{\alpha}(t)=\pi_1+
[p_1^{\alpha}(0)-\pi_1]e^{-\lambda t}.
\end{equation}
Since $T_H>T_C>T_b$, one has
$p_1^H(t)>p_1^C(t)>\pi_1$ at every finite time. The derivative of binary
relative entropy is
\begin{equation}
\frac{\mathrm dD(p\Vert\pi_1)}{\mathrm dp}
=\ln\frac{p(1-\pi_1)}{\pi_1(1-p)}>0
\quad\text{when }p>\pi_1.
\end{equation}
Consequently $D_H(t)>D_C(t)$ and there is no crossing. The numerical
control agrees with this exact conclusion. The theorem's next-distinct-rate
assumption also explains why its two-sector formula is not applicable
to this one-mode case.

\section{Certified crossings and a continuous preparation region}
\label{sec:certified_robustness}

The general bounds are analytical inequalities. Their evaluation becomes
a numerical certificate only when the input errors and sign comparisons
are also enclosed. Table~\ref{tab:certified-models} summarizes the claims
verified independently with Arb ball arithmetic; the corresponding
proof algorithms are given in Appendix~\ref{app:certification}.
The selected network and diffusion certificates use direct semigroup
propagation for their first-meeting intervals. Separate validated
eigenvalue and eigenvector enclosures establish their positive slow-sector
margins and upper formulas. The eight-state partial-spectrum endpoint
$t_M<1.55163$ is also validated, including its omitted-tail cutoff.

To demonstrate a full region of thermal preparations, consider the
exact three-state benchmark
\begin{equation}
E=(0,2/5,3),\qquad
(\gamma_{01},\gamma_{02},\gamma_{12})=(1,23/5,3/25),
\label{eq:robust_benchmark}
\end{equation}
with symmetric prefactors, $T_b=1$, and the rate law in
Eq.~\eqref{eq:network-rates}. Both preparations always experience this
same fixed generator. At $(T_H,T_C)=(3,3/2)$, the hot slow weight is
strictly positive and the first meeting lies in
$(0.17022,0.17024)$. This reference point is not an exactly strong
preparation.

More substantially, every independent temperature pair in the closed
rectangle
\begin{equation}
2.9\le T_H\le3.1,\qquad 1.45\le T_C\le1.55
\label{eq:certified_temperature_box}
\end{equation}
has its first meeting in
\begin{equation}
\boxed{0.161<t_M<0.179.}
\label{eq:temperature_box_crossing}
\end{equation}
Throughout this entire rectangle, interval evaluation gives
$w_H>3.2401\times10^{-6}$ and $\Gamma>1.8624\times10^{-4}$.
Thus no hot preparation in the certified region cancels its slow mode.

\begin{figure}[t]
\centering
\includegraphics[width=\columnwidth]{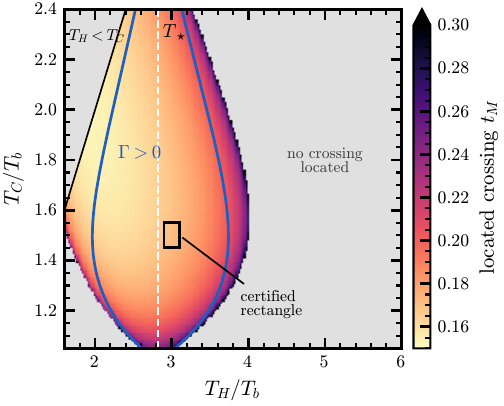}
\caption{Located crossing time of the three-state benchmark over the plane
of preparation temperatures (double precision; gray: $T_H<T_C$ or no
crossing located). The blue contour is $\Gamma=0$, with $\Gamma>0$ inside; the
dashed line marks $T_\star$, where the hot slow weight vanishes. The black
box is the rectangle of Eq.~\eqref{eq:certified_temperature_box}, in which
every pair is certified by interval arithmetic to have its first crossing
in $(0.161,0.179)$.}
\label{fig:benchmark_map}
\end{figure}

Figure~\ref{fig:benchmark_map} places this rectangle in the full plane of
preparation temperatures. In double-precision evaluations on a
$177\times136$ grid, crossings are located only in a lobe around the
strong temperature $T_\star\simeq2.823\,T_b$, where the hot slow weight is
small. The margin is positive in the interior of that lobe, covering
about $75\%$ of the grid points with a located crossing, and every grid
point with $\Gamma>0$ has a located crossing, as Theorem~1 requires. The
certified rectangle lies well inside the $\Gamma>0$ region; crossings
therefore do not depend on tuning $T_H$ to $T_\star$.

The interior of this rectangle is an open set of preparation pairs with
the same crossing property. The statement includes intermediate
temperatures and the boundary; it is not inferred from a point grid.

The certificate covers the hot and cold temperature axes by adjacent
rational intervals, encloses all Gibbs probabilities and modal
coefficients over each interval, and then applies entropy envelopes to
their combined ranges. A bootstrap excludes every meeting up to $0.161$,
while a uniform negative upper comparison at $0.179$ supplies the other
endpoint. The construction keeps the uncertainty correlations within
each preparation enclosed and allows the two temperatures to vary
independently. It does not prove that this is the largest permitted
temperature region or that all trajectories have only one meeting.

An arbitrary probability perturbation has a different uncertainty set.
At the central pair $(3,3/2)$, independent normalized initial errors
$\|\delta p_H\|_1,\|\delta p_C\|_1\le10^{-6}$ preserve a first
meeting in $(0.169,0.172)$. For all such errors the likelihood ratios
remain in $[0.7,6]$, giving
$C_H=C_C=\tfrac12\ln(6/0.7)$. The resulting total entropy-error budget
is less than $2.149\times10^{-6}$, below the separately certified sign
margins. This is a direct application of Theorem~2. The larger thermal
rectangle is obtained by enclosing its specific Gibbs family rather
than treating every direction in probability space as an allowed error.

\begin{table*}[t]
\centering
\caption{First-meeting statements established by interval arithmetic for
the exact inputs specified in the text and supplement. Each lower
endpoint is supported by an exclusion on the whole preceding time
interval. None of these rows asserts global uniqueness, certification
of an entire sampled network ensemble, or a continuum limit.}
\label{tab:certified-models}
\begin{tabular}{@{}p{0.47\textwidth}p{0.19\textwidth}p{0.27\textwidth}@{}}
\toprule
Model and preparations & Certified first meeting & Additional scope\\
\midrule
Three-state benchmark, $(T_H,T_C)=(3,1.5)$ & $(0.17022,0.17024)$ & Hot slow weight strictly positive\\
Same benchmark, temperature rectangle \eqref{eq:certified_temperature_box} & $(0.161,0.179)$ & Every pair in the closed rectangle\\
Same benchmark, each preparation perturbed in $1$-norm by at most $10^{-6}$ & $(0.169,0.172)$ & Fixed final generator\\
Selected eight-state complete graph & $(1.26,1.28)$ & Exact decimal prefactors\\
Reflecting diffusion, $N=64$ cells & $(3,3.2)$ & Specified finite discretization\\
\bottomrule
\end{tabular}
\end{table*}

\section{Physical interpretation, scope, and limitations}
\label{sec:discussion}

The crossing condition compares how the two preparations distribute their
initial departures from equilibrium across slow and fast relaxation
patterns. A hot state may have a larger total departure but a smaller
entropy-weighted slow contribution. The excess fast contribution then
decays sufficiently quickly for the ordering to reverse. The positive
margin $\Gamma$ supplies a quantitative allowance for a residual hot slow
weight; the separation $\lambda_f-\lambda_s$ sets the time needed for the
remaining fast contribution to become smaller than that margin.

This interpretation distinguishes ordinary crossing from strong
asymptotic acceleration. If both slow weights are nonzero, both relative
entropies eventually decay with the same exponent, even though their
amplitudes and threshold times differ. The strong effect requires
cancellation of the entire slowest active eigenspace. Finite-time sign
margins are therefore useful when exact cancellation is unavailable or
unnecessary for the operational comparison.

Generality of the theorem does not mean uniform sharpness. Its
assumptions refer to any finite size, but its constants depend on the
initial likelihood-ratio ranges and spectral separation. Small
stationary probabilities or a weak separation can make the bound loose.
The network study displays this directly: among the 45 positive-margin
cases, the ratio of the simple upper formula to the located crossing
ranges from about 1.21 to 30.19, with median 2.15. Retaining more spectral
information can improve a particular bound substantially, as the
eight-state example demonstrates, but no dimension-independent accuracy
guarantee follows.

The partial-spectrum construction is an information bound. To use it as
a numerical certificate, one needs enclosures for the retained weights
and rates, the total initial squared deviation, and a valid lower bound
on the omitted spectrum. The demonstration here computes the full
spectrum once and then evaluates bounds with fewer retained modes. For
the selected eight-state model, interval arithmetic validates this
construction and the endpoint $t_M<1.55163$. This does not establish the
computational complexity or speedup of a large matrix eigensolver.
Certified iterative spectral methods and sharper tail estimates would
be useful subsequent developments.

The diffusion example connects the finite-state argument to an explicit
Brownian relaxation model. The grid generator preserves detailed balance
and probability exactly at the analytical level, and the observed
crossing stabilizes under refinement. The interval certificate applies
to a specified finite grid. A continuum crossing theorem with a
numerical error enclosure would additionally require a discretization
estimate for both distributions or their entropy difference. On an
unbounded domain, hotter Gibbs preparations can also have unbounded
likelihood ratios, requiring tail control beyond the present finite-range
comparison.

Several extensions have distinct mathematical requirements. The scalar
entropy comparison remains valid for a fixed nonreversible stationary
generator, and the chi-squared divergence remains contractive, but the
orthogonal sum of modal squares generally does not survive. The exact
nonadiabatic entropy balance still holds at a fixed stationary reference
\cite{esposito2010}; it does not justify replacing reversible decay rates
by arbitrary complex eigenvalues. Time-dependent generators and changing
reference states also introduce additional terms. Finally, an entropy
crossing alone need not imply a crossing of mean energy, temperature, or
every monotone distance \cite{vu2025,summer2026}. The present analysis
therefore makes no statement about which water sample freezes first.

The results identify a concrete combination of established ingredients
and derived applications. The spectral Mpemba mechanism, stochastic
contraction, entropy dissipation, and the static divergence comparison
are prior results. Their use here yields finite-amplitude crossing
enclosures, a controlled omitted-spectrum contribution, quantitative
preparation tolerances, and reproducible tests with explicitly separated
certification levels. No claim of priority for all finite-time Mpemba
criteria, or of a universal optimization law, is needed for these
statements.

\section{Conclusions}
\label{sec:conclusions}

A finite-time relative-entropy crossing can be guaranteed without exact
hot slow-mode cancellation. For a reversible finite-state system, a
positive slow-sector margin gives an explicit upper meeting-time bound,
with the complete slow eigenspace included in any degeneracy. Exact
entropy envelopes retain additional modes, and a bounded omitted tail
makes partial spectral information usable without silently discarding
its error. Probability contraction and an entropy-gradient estimate
convert sign margins into preparation tolerances.

The 2000-network study and the reflecting diffusion example show how
these results operate beyond a specially canceled three-state
preparation. The simple condition detects 45 of 86 numerically located
network crossings in the declared sample and can be conservative.
Independent interval arithmetic verifies selected spectral bounds,
first crossings, a partial-spectrum improvement, and a continuous
temperature rectangle in an exactly specified benchmark.
Together these results supply a general finite-state framework with
explicit hypotheses and model-dependent numerical quality, while
separating ordinary crossings from strong asymptotic acceleration.

\section*{Data and code availability}

The Supplemental Material contains all $2000$ network draws, all $180$
diffusion-search outcomes, the mesh-refinement and temperature-survey
data, exact inputs for the certified examples, the numerical driver
(\texttt{run\_generalization.py}), the figure script, and the
interval-arithmetic program (\texttt{certify\_generalization.py}), with a
README giving commands, software versions, and the scope of each
certificate. Rounded values in tables and figures do not replace the
exact model definitions used by the certificates.

\appendix

\section{Entropy balance, canonical ordering, and reversible modes}
\label{app:foundations}

\subsection{Thermodynamic meaning and canonical initial ordering}

For $\pi_i=e^{-E_i/T_b}/Z(T_b)$,
\begin{align}
T_bD(p\Vert\pi)
&=T_b\sum_i p_i\ln p_i+\sum_i p_iE_i+T_b\ln Z(T_b)\\
&=\mathcal F[p]-\mathcal F[\pi],
\end{align}
because $\mathcal F[\pi]=-T_b\ln Z(T_b)$. This identifies the
dimensionless entropy distance with excess free energy when the
reference is Gibbsian.

For an initial Gibbs state at inverse temperature $\beta_0=1/T_0$,
write $U(\beta_0)=\sum_i E_i e^{-\beta_0E_i}/Z(\beta_0)$ and
$\beta_b=1/T_b$. Direct substitution gives
\begin{equation}
D_0(\beta_0)=(\beta_b-\beta_0)U(\beta_0)
+\ln\frac{Z(\beta_b)}{Z(\beta_0)}.
\end{equation}
Using $\partial_{\beta_0}\ln Z=-U$ and
$\partial_{\beta_0}U=-\operatorname{Var}_{T_0}(E)$, differentiation
cancels the two terms proportional to $U$ and yields
\begin{equation}
\frac{\partial D_0}{\partial T_0}
=\frac{\beta_b-\beta_0}{T_0^2}\operatorname{Var}_{T_0}(E)>0
\quad(T_0>T_b).
\end{equation}
The variance is strictly positive for nonconstant energies. Thus the
canonical cooling protocol starts with $D_H(0)>D_C(0)>0$.

\subsection{The entropy-production identity}

Differentiate $D=\sum_i p_i\ln(p_i/\pi_i)$ with fixed $\pi$.
Normalization removes the constant derivative term, giving
\begin{equation}
\dot D=\sum_i\dot p_i\ln\frac{p_i}{\pi_i}.
\end{equation}
Since $\dot p_i=-\sum_jJ_{ij}$ and $J_{ij}=-J_{ji}$,
\begin{equation}
\dot D=-\frac12\sum_{i,j}J_{ij}
\left[\ln\frac{p_i}{\pi_i}-\ln\frac{p_j}{\pi_j}\right].
\end{equation}
Detailed balance identifies the logarithmic difference with
$\ln(k_{ij}p_i/k_{ji}p_j)$. Therefore $\dot D=-\dot e_p$.
Each paired contribution to $\dot e_p$ has the form
$(a-b)\ln(a/b)\ge0$. Irreducibility gives equality only at $p=\pi$.
Integrating $-\mathrm d\ln D/\mathrm dt=\dot e_p/D$ proves
Eq.~\eqref{eq:clock}.

\subsection{Self-adjointness and orthogonal amplitudes}

Put $\Pi=\operatorname{diag}(\pi)$. Dividing the master equation by
$\pi_i$ and using detailed balance gives
\begin{equation}
\dot f_i=\sum_{j\ne i}k_{ij}(f_j-f_i),\qquad
W=\Pi L\Pi^{-1}.
\end{equation}
For real vectors $u,v$,
\begin{align}
\langle u,Lv\rangle_\pi
&=\sum_{i,j}\pi_i k_{ij}u_i(v_j-v_i)\\
&=-\frac12\sum_{i,j}\pi_i k_{ij}(u_i-u_j)(v_i-v_j)\\
&=\langle Lu,v\rangle_\pi.
\end{align}
The second line follows by interchanging $i,j$ in one copy of the sum
and averaging, with $\pi_i k_{ij}=\pi_j k_{ji}$. Thus $L$ is
self-adjoint and nonpositive. Its constant null eigenfunction is
$\phi_1=1$; irreducibility makes that null space one-dimensional.
The finite-dimensional spectral theorem gives the remaining
orthonormal eigenfunctions in Eq.~\eqref{eq:spectrum}.
Since $\langle f(0),1\rangle_\pi=0$, no stationary coefficient appears.
Taking inner products of the mode expansion with $\phi_n$ gives
Eq.~\eqref{eq:fexp}.

Squaring that expansion and averaging makes the cross terms vanish:
\begin{align}
Q_\alpha(t)
&=\sum_{m,n\ge2}b_{m,\alpha}b_{n,\alpha}
e^{-(\lambda_m+\lambda_n)t}\langle\phi_m,\phi_n\rangle_\pi\\
&=\sum_{n\ge2}b_{n,\alpha}^2e^{-2\lambda_nt}.
\end{align}
For numerical diagonalization, the similar matrix
$\Pi^{-1/2}W\Pi^{1/2}$ is real symmetric. An orthonormal eigenvector
$z_n$ of this matrix gives $\phi_{n,i}=z_{n,i}/\sqrt{\pi_i}$.

\subsection{The quadratic limit and the asymptotic weight criterion}

For $|x|<1$, the scalar entropy function has the convergent expansion
\begin{equation}
g(x)=\frac{x^2}{2}-\frac{x^3}{6}+\frac{x^4}{12}-\cdots.
\end{equation}
If $\eta=\|f\|_\infty<1$, its remainder after the quadratic term is
bounded componentwise by a constant times $\eta f_i^2$, with a constant
bounded on each fixed range $\eta\le\eta_0<1$. Summation gives
$D=Q/2+O(\eta Q)$ and proves Eq.~\eqref{eq:quadratic_limit}.

If the cold preparation has $w_C>0$, then
$Q_C(t)=w_Ce^{-2\lambda_st}+o(e^{-2\lambda_st})$.
For the hot preparation the corresponding leading coefficient is $w_H$,
with a vanishing leading term if $w_H=0$. Both relative deviations tend
to zero, so
\begin{equation}
\lim_{t\to\infty}\frac{D_H(t)}{D_C(t)}=\frac{w_H}{w_C}.
\end{equation}
Thus $w_H<w_C$ and the opposite initial ordering imply an eventual
meeting. This asymptotic statement alone supplies no finite upper time.
Since $m_H(0)\le0\le M_C(0)$ and $F(0)=1/2$, one has
$c_H\ge1/2$ and $a_C\le1/2$. Hence $\Gamma>0$ implies
$w_H/w_C<a_C/c_H\le1$, explaining why Theorem 1 is a stronger
sufficient condition.

\section{Finite-amplitude comparison and the crossing theorems}
\label{app:general_proofs}

\subsection{The scalar bound and Markov range contraction}

Taylor's integral formula, or two direct integrations of
$g''(x)=1/(1+x)$, gives
\begin{equation}
g(x)=\int_0^x\frac{x-y}{1+y}\,\mathrm dy
=x^2\int_0^1\frac{1-u}{1+ux}\,\mathrm du.
\end{equation}
For $x>-1$ the denominator is positive for $0\le u\le1$.
This proves positivity of $F$ and permits differentiation under the
integral to obtain Eq.~\eqref{eq:Fmonotone}. Also
$\sum_i\pi_i f_i=0$, so
\begin{equation}
D(p\Vert\pi)=\sum_i\pi_i[(1+f_i)\ln(1+f_i)-f_i]
=\sum_i\pi_i f_i^2F(f_i).
\end{equation}

Let $P(u)=e^{Wu}$. Its entries are nonnegative and its columns sum to
one. Stationarity gives $P(u)\pi=\pi$, so the ratio kernel
$K_{ij}(u)=P_{ij}(u)\pi_j/\pi_i$ has row sums one. Therefore
\begin{equation}
m_\alpha(s)\le\sum_jK_{ij}(t-s)f_j^\alpha(s)
\le M_\alpha(s),\qquad t\ge s.
\end{equation}
Because $F$ decreases, every term in the exact entropy sum lies between
$\pi_i(f_i^\alpha)^2F(M_\alpha(s))$ and
$\pi_i(f_i^\alpha)^2F(m_\alpha(s))$. Summing proves
Eq.~\eqref{eq:envelopes} without a small-amplitude assumption.

\subsection{A whole slow eigenspace and the explicit upper time}

Let $\mathcal S=\{n\ge2:\lambda_n=\lambda_s\}$ and
$\mathcal F=\{n\ge2:\lambda_n>\lambda_s\}$. Under Theorem 1,
$\mathcal F$ is nonempty and $\lambda_f$ is its smallest rate.
The weights are $w_\alpha=\sum_{n\in\mathcal S}b_{n,\alpha}^2$ and
$q_H=\sum_{n\in\mathcal F}b_{n,H}^2$. An orthogonal rotation inside
$\mathcal S$ preserves this sum of squares, so a degenerate sector is
handled without selecting a special basis vector.

Since every spectral weight is nonnegative,
\begin{align}
Q_H(t)
&=w_He^{-2\lambda_st}
+\sum_{n\in\mathcal F}b_{n,H}^2e^{-2\lambda_nt}\\
&\le w_He^{-2\lambda_st}+q_He^{-2\lambda_ft},
\end{align}
and dropping the faster cold terms gives
$Q_C(t)\ge w_Ce^{-2\lambda_st}$. Inserting these two inequalities
into $d(t)\le c_HQ_H(t)-a_CQ_C(t)$ proves
Eq.~\eqref{eq:general_difference}.

At zero time $d(0)>0$, while that upper expression equals
$c_Hq_H-\Gamma$. Thus $c_Hq_H>\Gamma>0$; in particular $q_H>0$
and $w_C>0$. Solving
\begin{equation}
c_Hq_He^{-2(\lambda_f-\lambda_s)t}=\Gamma
\end{equation}
gives the positive time in Eq.~\eqref{eq:general_upper}. The comparison
is nonpositive at that time and strictly negative afterward.
The continuous function $d$ starts positive, so it has a first zero in
$(0,T_{\rm up}]$ and a strict reversed ordering for every
$t>T_{\rm up}$. This proves Theorem 1 and does not exclude an earlier
tangency or additional earlier crossings.

If $d(s)>0$, the proof can be restarted using the distributions at $s$.
The weights must be those of $f^\alpha(s)$, and the exponential variable
is $t-s$. A positive restarted margin gives an absolute upper time
$s+T_{\rm up}^{(s)}$. This also bounds the original first meeting even
if there was an earlier meeting and recrossing. If $d(s)\le0$,
continuity already gives $t_M\le s$; a negative restarted logarithm
must not be used to move the bound into the past of its reference time.

\subsection{All-mode signs and the omitted spectral tail}

For fixed $s$, Eq.~\eqref{eq:envelopes} bounds each entropy separately.
Subtracting the upper cold bound from the lower hot bound gives
$G_-(t;s)\le d(t)$. Subtracting the lower cold bound from the upper hot
bound gives $d(t)\le G_+(t;s)$. This proves
Eq.~\eqref{eq:general_envelopes}. Positivity of the lower bound on an
entire earlier interval excludes zeros there. Nonpositivity of the upper
bound at a later endpoint guarantees that at least one zero has already
occurred, since the initial difference was positive. Neither inference
requires monotonicity of the bounding exponential sums.

For omitted modes,
\begin{equation}
R_\alpha(t)=\sum_{n=m+1}^N b_{n,\alpha}^2e^{-2\lambda_nt}.
\end{equation}
Every term is nonnegative and, by the rate lower bound, is no larger
than $b_{n,\alpha}^2e^{-2\underline\lambda_{\rm cut}t}$.
Summation proves Eq.~\eqref{eq:general_tail}. If uncertain spectral
inputs are enclosed as in Eq.~\eqref{eq:partial_spectrum_error_bounds},
monotonicity of $e^{-2\lambda t}$ for $t\ge0$ gives
\begin{equation}
\underline w_{n,\alpha}e^{-2\overline\lambda_nt}
\le b_{n,\alpha}^2e^{-2\lambda_nt}
\le\overline w_{n,\alpha}e^{-2\underline\lambda_nt}.
\end{equation}
Adding these retained terms and the omitted upper tail proves the stated
norm bounds. A fully explicit entropy-difference enclosure is therefore
\begin{equation}
\begin{aligned}
d(t)&\ge a_H(s)Q_H^-(t)-c_C(s)Q_C^+(t),\\
d(t)&\le c_H(s)Q_H^+(t)-a_C(s)Q_C^-(t),
\qquad t\ge s.
\end{aligned}
\label{eq:partial_difference_bounds}
\end{equation}
If the factors themselves are intervals, their nonnegative upper or
lower endpoints must be chosen according to these inequality directions.
The formulas require valid interval inputs, not merely approximate
spectral estimates.

\section{Proofs of preparation and parameter tolerance}
\label{app:robust_proofs}

\subsection{Probability contraction and the centered entropy gradient}

For the common stochastic semigroup $P(t)=e^{Wt}$,
\begin{align}
\|\widetilde p(t)-p(t)\|_1
&=\sum_i\left|\sum_jP_{ij}(t)
[\widetilde p_j(0)-p_j(0)]\right|\\
&\le\sum_{i,j}P_{ij}(t)|\widetilde p_j(0)-p_j(0)|\\
&=\|\widetilde p(0)-p(0)\|_1.
\end{align}
The last equality uses the unit column sums. The ratio kernel separately
preserves the common initial likelihood-ratio range for both trajectories.

At a fixed time put $\delta p=\widetilde p-p$ and
$p_u=p+u\delta p$, $0\le u\le1$. Every point on this segment is a
normalized probability distribution with ratios in $[h_-,h_+]$.
Differentiating the entropy along the segment and integrating gives
\begin{equation}
\begin{aligned}
D(\widetilde p\Vert\pi)-D(p\Vert\pi)
&=\int_0^1\sum_i\delta p_i
\left[\ln\frac{p_{u,i}}{\pi_i}+1\right]\,\mathrm du.
\end{aligned}
\end{equation}
Since $\sum_i\delta p_i=0$, subtract the constant
$1+\tfrac12(\ln h_-+\ln h_+)$ inside the brackets without changing
the integral. The remaining factor has magnitude at most
$C=\tfrac12\ln(h_+/h_-)$. Therefore
\begin{equation}
|D(\widetilde p\Vert\pi)-D(p\Vert\pi)|
\le C\|\widetilde p-p\|_1.
\end{equation}
Combining this with probability contraction proves each entropy-error
bound in Eq.~\eqref{eq:entropy_error}; the triangle inequality gives the
bound on $\widetilde d-d$.

Under Theorem 2's error budget, for every $0\le t\le L$ one has
$\widetilde d(t)\ge\eta_L-\delta_D>0$, while
$\widetilde d(U)\le-\eta_U+\delta_D<0$.
Hence no earlier zero exists and continuity places a first zero in
$(L,U)$. This proves Theorem 2.

To justify the component error used in Eq.~\eqref{eq:range_from_budget},
decompose a normalized difference into positive and negative parts.
Their total masses are equal because $\sum_i\delta p_i=0$, and their
sum is $\|\delta p\|_1\le\epsilon$. Each mass is therefore at most
$\epsilon/2$, implying $|\delta p_i|\le\epsilon/2$ for every state.
The stated ratio endpoints contain every allowed preparation if their
lower endpoint is positive.

\subsection{Gibbs temperature sensitivity}

For $p_i(T)=e^{-E_i/T}/Z(T)$,
\begin{equation}
\frac{\partial\ln p_i}{\partial T}
=\frac{E_i}{T^2}-\frac{\partial\ln Z}{\partial T}
=\frac{E_i-U(T)}{T^2}.
\end{equation}
Multiplication by $p_i(T)$ proves Eq.~\eqref{eq:Gibbs_derivative} and
summation of absolute values gives
\begin{equation}
\left\|\frac{\partial p(T)}{\partial T}\right\|_1
=\frac{\langle|E-U|\rangle_T}{T^2}
\le\frac{\sqrt{\operatorname{Var}_T(E)}}{T^2}.
\end{equation}
The last step is Cauchy--Schwarz. To bound the variance, let
$a=E_{\min}$, $b=E_{\max}$ and use $(E-a)(b-E)\ge0$.
Taking expectations gives
$\langle E^2\rangle\le(a+b)U-ab$, hence
\begin{equation}
\operatorname{Var}(E)\le(U-a)(b-U)\le\frac{(b-a)^2}{4}.
\end{equation}
Thus the norm of the temperature derivative is at most
$(E_{\max}-E_{\min})/(2T^2)$. Integrating along the interval between
$T$ and $\widetilde T$, whose temperatures are all at least
$T_{\min}>0$, gives Eq.~\eqref{eq:temperature_error}.

\subsection{A change of rates and stationary reference}
\label{app:rate_perturbations}

Let $\Delta W=\widetilde W-W$. Differentiating
$e^{\widetilde W(t-s)}e^{Ws}$ with respect to $s$ and integrating yields
\begin{equation}
e^{\widetilde Wt}-e^{Wt}
=\int_0^t e^{\widetilde W(t-s)}\Delta W e^{Ws}\,\mathrm ds.
\end{equation}
Each stochastic semigroup has induced $1$-norm equal to one. Splitting
the probability difference into its initial-state and generator parts,
\begin{align}
\widetilde p(t)-p(t)
&=e^{\widetilde Wt}[\widetilde p(0)-p(0)]\\
&\quad+(e^{\widetilde Wt}-e^{Wt})p(0),
\end{align}
and using $\|p(0)\|_1=1$ proves
Eq.~\eqref{eq:generator_error}.

The centered-gradient bound can now compare $\widetilde p(t)$ and
$p(t)$ relative to the \emph{same nominal} $\pi$. Its ratio interval
must be valid for $0\le t\le U$, since the perturbed generator need
not preserve a range relative to the old reference. A conservative
construction is available from Eq.~\eqref{eq:generator_error}. Put
$e_\alpha=\epsilon_\alpha+U\delta_W$ and
$\pi_{\min}=\min_i\pi_i$. The nominal range is preserved, and every
component difference is at most $e_\alpha/2$. Thus
\begin{equation}
\begin{aligned}
h_{\alpha,-}&=1+m_\alpha(0)-\frac{e_\alpha}{2\pi_{\min}},\\
h_{\alpha,+}&=1+M_\alpha(0)+\frac{e_\alpha}{2\pi_{\min}}
\end{aligned}
\end{equation}
encloses both distributions' ratios to $\pi$ whenever
$h_{\alpha,-}>0$. More detailed component bounds can improve it.

The stationary-reference correction is exact:
\begin{equation}
D(\widetilde p\Vert\widetilde\pi)
-D(\widetilde p\Vert\pi)
=\sum_i\widetilde p_i\ln\frac{\pi_i}{\widetilde\pi_i}.
\end{equation}
Its absolute value is at most
$\delta_\pi=\max_i|\ln(\pi_i/\widetilde\pi_i)|$.
Combining this with the distribution error for each preparation and
adding the two bounds proves Eq.~\eqref{eq:rate_error_budget}.
The perturbed stationary distribution must itself be computed or bounded
for the specified rate family. The resulting statement is finite-time:
the linear-in-$U$ Duhamel estimate need not remain useful on arbitrarily
long intervals.

\subsection{Differentiating a transverse meeting}

For a simple root $d(t_*(\theta),\theta)=0$, the chain rule gives
\begin{equation}
0=\partial_t d\,\partial_\theta t_*+\partial_\theta d.
\end{equation}
At the root the two positive entropies equal $D_*$, and the entropy
balance gives
\begin{equation}
\partial_t d=-\dot e_p^{H}+\dot e_p^{C}=-D_*(r_H-r_C).
\end{equation}
Substitution proves Eq.~\eqref{eq:crossing_sensitivity}. The nonzero
derivative condition is essential to the local implicit-function
argument. It establishes neither global uniqueness nor absence of
earlier roots; those questions require separate sign information.

\section{A negative control and the scope beyond detailed balance}
\label{app:scope}

\subsection{Two-state canonical cooling cannot reverse entropy order}

Let $E_1>E_0$ and denote the excited-state probability by $u(T)$.
For a Gibbs preparation,
\begin{equation}
u(T)=\frac{1}{1+e^{(E_1-E_0)/T}},
\end{equation}
which is strictly increasing in $T$. Write $u_b=u(T_b)$.
The two-state master equation has one nonzero relaxation rate
$\lambda_2=k_{01}+k_{10}$ and exact solution
\begin{equation}
u_\alpha(t)=u_b+[u(T_\alpha)-u_b]e^{-\lambda_2t}.
\end{equation}
Thus $T_H>T_C>T_b$ implies
$u_H(t)>u_C(t)>u_b$ for every finite time. Binary relative entropy is
\begin{equation}
D(u\Vert u_b)=u\ln\frac{u}{u_b}
+(1-u)\ln\frac{1-u}{1-u_b},
\end{equation}
and its derivative is
\begin{equation}
\frac{\mathrm dD}{\mathrm du}
=\ln\frac{u(1-u_b)}{u_b(1-u)}>0\qquad (u>u_b).
\end{equation}
Consequently $D_H(t)>D_C(t)$ at every finite time. This is a negative
control for the stated canonical cooling protocol, not a claim about
arbitrary binary preparations on opposite sides of equilibrium.

\subsection{What survives without reversibility}

For any fixed irreducible Markov generator with positive stationary
distribution $\pi$, the ratio kernel in Eq.~\eqref{eq:ratio_kernel}
still has nonnegative entries and unit row sums. Therefore the scalar
entropy envelopes and likelihood-ratio range contraction remain valid.
The exact squared deviation can be written as
\begin{equation}
Q_\alpha(t)=\left\|\Pi^{-1/2}e^{Wt}
[p^\alpha(0)-\pi]\right\|_2^2.
\end{equation}
Without detailed balance it cannot generally be replaced by a sum of
independent squared eigenmode amplitudes, because the modes can be
nonorthogonal or complex.

Nevertheless $Q$ is nonincreasing. Probability conservation gives
$\sum_i\pi_iK_{ij}=\pi_j$, and Jensen's inequality yields
\begin{align}
\sum_i\pi_i\left(\sum_jK_{ij}f_j\right)^2
&\le\sum_{i,j}\pi_iK_{ij}f_j^2\\
&=\sum_j\pi_jf_j^2.
\end{align}
This proves chi-squared contraction without reversibility. Nonorthogonal
spectral representations do not imply transient amplification of this
stationary-weighted divergence.

For completeness, set $h_i=p_i/\pi_i$. Differentiating relative entropy
with a fixed stationary reference gives
\begin{equation}
-\dot D=\sum_{i,j}\pi_i k_{ij}h_i\ln\frac{h_i}{h_j}.
\end{equation}
Stationarity cancels
$\sum_{i,j}\pi_i k_{ij}(h_i-h_j)$, so
\begin{equation}
-\dot D=\sum_{i,j}\pi_i k_{ij}
\left[h_i\ln\frac{h_i}{h_j}-h_i+h_j\right]\ge0.
\end{equation}
This is the nonadiabatic entropy-production rate. With bidirectional
allowed transitions, finite total entropy production can also contain
a housekeeping contribution \cite{esposito2010}. A time-dependent reference adds
$-\sum_i p_i\partial_t\ln\pi_i$ to $\dot D$, so the fixed-reference
clock must not be applied unchanged to that situation.

Finally, every finite detailed-balance discretization of a diffusion
model falls within the finite-state results. A continuum theorem would
also need the required operator and likelihood-ratio bounds in the
continuum setting. On an unbounded domain a hotter Gibbs distribution
can have an unbounded ratio to the final Gibbs density, making the
present global lower entropy factor trivial. Mesh refinement supplies
numerical evidence of convergence; it does not by itself certify a
continuum crossing interval.

\section{Reproducible numerical methods and consistency checks}
\label{sec:numerical-methods}

\subsection{Generator assembly and random streams}

The generator acts on column probability vectors. From jump rates
$k_{ij}$ for $i\to j$, the off-diagonal entry is $W_{ji}=k_{ij}$ and the
diagonal entry is $W_{ii}=-\sum_{j\ne i}k_{ij}$. Thus each column sums to
zero. For a reversible model, diagonalize the symmetric positive
semidefinite matrix
\begin{equation}
S=-\Pi^{-1/2}W\Pi^{1/2},\qquad
\Pi=\operatorname{diag}(\pi_1,\ldots,\pi_N).
\end{equation}
If $u_n$ is a Euclidean-normalized eigenvector of $S$, then
$\phi_n=\Pi^{-1/2}u_n$ is normalized in the weighted inner product used
in the theory. The nonstationary coefficient can be computed without
subtracting large entropy values:
\begin{equation}
b_{n,\alpha}=\sum_i\phi_{n,i}[p_i^\alpha(0)-\pi_i].
\end{equation}
The stationary eigenvector is omitted analytically. For the rates in
Eq.~\eqref{eq:network-rates}, the off-diagonal entries of $S$ are simply
$-\gamma_{ij}$, avoiding a numerical symmetrization mismatch. Random
continuous prefactors give nondegenerate computed slow rates in the
reported samples; the theorem itself groups a degenerate slow eigenspace.

The network iteration order is complete, sparse connected, ring,
birth--death; within each family the sizes are $4,8,16,32,64$, and within
each size the draw index runs from $0$ to $99$. Every case uses a fresh
NumPy \texttt{default\_rng} initialized by
\begin{equation}
\begin{gathered}
\texttt{SeedSequence}(20261024,\\
\texttt{spawn\_key}=(f,N,d)),
\end{gathered}
\end{equation}
where $f$ is the zero-based family index and $d$ the zero-based draw
index. The energy permutation is drawn first. For a sparse graph, edge
Bernoulli variables are drawn next over upper-triangular pairs in NumPy
order, and the ring backbone is then imposed by a logical union. Finally
the log-prefactors are drawn on allowed pairs in the same order.
Specifying this order makes every realization independently reproducible.
The energy permutation and spawn key are retained for every draw.

The selected network's interval calculation uses the rational energy
grid in Eq.~\eqref{eq:selected-network-energy} and the supplied decimal
prefactors as exact inputs. Ordinary numerical calculations use their
floating-point representations. The rigorous input file, rather than
an assumed exact meaning of a pseudorandom exponential, defines the
reference model for that certificate.

\subsection{Stable entropy and crossing search}

For nonstationary modal propagation,
\begin{equation}
f_i^\alpha(t)=\sum_{n=2}^N b_{n,\alpha}\phi_{n,i}e^{-\lambda_nt},
\end{equation}
we evaluate
\begin{equation}
D_\alpha(t)=\sum_i\pi_i h(f_i^\alpha(t)),\qquad
h(x)=(1+x)\ln(1+x)-x.
\end{equation}
The subtracted linear term sums to zero exactly in the mathematical
model. For $|x|<10^{-3}$, direct evaluation can lose relative accuracy,
so the implementation instead sums
\begin{equation}
h(x)=\sum_{n=2}^{18}\frac{(-1)^n x^n}{n(n-1)}
+\mathcal R_{19}(x).
\end{equation}
The absolute remainder obeys
\begin{equation}
|\mathcal R_{19}(x)|
\le\frac{|x|^{19}}{19\cdot18(1-|x|)},\qquad |x|<1.
\end{equation}
The value of $F(x)=h(x)/x^2$ is evaluated using the corresponding series
near zero, with $F(0)=1/2$. All initial likelihood ratios are strictly
positive. No small-deviation approximation replaces the entropy curves.

For each numerical model the time search uses $\tau=\lambda_s t=0$
and $500$ logarithmically spaced points from $10^{-6}$ to $64$. When
available, the slow-sector upper endpoint is included if it is at most
$\tau=250$. In the reported network study all added endpoints are below
$64$. To locate an interchange, the code searches for adjacent points
where $\ln(D_H/D_C)$ changes from positive to nonpositive, and refines
that bracket with Brent's method. The absolute time tolerance is
$10^{-13}$ and the relative tolerance $2\times10^{-14}$; these are solver
settings, not rigorous root-error bounds. The finite time grid cannot
exclude an earlier narrow crossing and recrossing. Accordingly, reported
roots are called located crossings unless an additional interval proof
establishes their first-crossing property.

The smallest absolute computed $\Gamma$ across the network draws is
$8.76\times10^{-9}$, and the smallest absolute computed difference
$|w_H-w_C|$ is $8.97\times10^{-8}$. Every nonstationary rate is positive
at working precision; the smallest is $2.5039\times10^{-5}$. Every sampled
slow rate is simple at working precision. More specifically,
\begin{align}
\min\frac{\lambda_f-\lambda_s}{\lambda_f}
&\simeq0.0075115,\\
\min\frac{\lambda_f-\lambda_s}{\lambda_{\max}}
&\simeq3.0244\times10^{-7}.
\end{align}
The driver rejects a slow gap smaller than $10^{-10}\lambda_{\max}$,
because it uses one slow eigenvector; a degenerate model must instead
group its whole slow eigenspace as required by the theorem. The largest
absolute computed stationary eigenvalue is $9.27\times10^{-14}$.
This quantity is a stationary-eigenvalue diagnostic, not a residual norm
for the entire eigensystem. These checks describe the numerical
calculation; they do not substitute for rigorous eigenvector or sign-error
enclosures for all $2000$ draws.

\subsection{Diffusion discretization and its flux limit}

Let $P_i(t)$ denote the probability in cell $i$ and write
$P_i=\rho_i\Delta x$. Define the rightward probability flux at the
interface by
\begin{equation}
J_{i+1/2}=k_{i,i+1}P_i-k_{i+1,i}P_{i+1}.
\end{equation}
The master equation is then
\begin{equation}
\dot\rho_i=-\frac{J_{i+1/2}-J_{i-1/2}}{\Delta x}.
\end{equation}
Using Eq.~\eqref{eq:diffusion-rates}, the interface flux becomes
\begin{equation}
J_{i+1/2}=
\frac{e^{-\Delta U_i/2}\rho_i-e^{\Delta U_i/2}\rho_{i+1}}{\Delta x},
\quad \Delta U_i=U_{i+1}-U_i.
\label{eq:discrete-flux}
\end{equation}
Expand about the midpoint $x_{i+1/2}$. For smooth $U$ and $\rho$,
\begin{align}
\Delta U_i&=\Delta x\,U'(x_{i+1/2})+O(\Delta x^3),\\
\rho_i&=\rho-\tfrac12\Delta x\,\rho'+O(\Delta x^2),\\
\rho_{i+1}&=\rho+\tfrac12\Delta x\,\rho'+O(\Delta x^2).
\end{align}
The even terms cancel between the two products in
Eq.~\eqref{eq:discrete-flux}. Thus
\begin{equation}
J_{i+1/2}=-\rho'-U'\rho+O(\Delta x^2).
\end{equation}
Taking the discrete divergence gives the stated Fokker--Planck equation
in the smooth-grid limit. At the two exterior interfaces, the flux is
set to zero by omitting external jumps. Also, when
$P_i\propto e^{-U_i}$, every interface flux vanishes exactly; equilibrium
is preserved by the discrete model.

The discretized stationary and initial Gibbs distributions are normalized
on equal-width cells, so the factor $\Delta x$ cancels. The symmetric
matrix is tridiagonal, with off-diagonal entries $-(\Delta x)^{-2}$ and
diagonal entries equal to the total escape rate. A symmetric tridiagonal
eigensolver is used for the mesh sequence
$N=32,64,128,256,512,1024$. The parameter search uses $N=128$ throughout,
before the selected potential is refined. The $81$-point temperature
survey also uses $N=128$ throughout. Search results, selection order,
and all mesh values are stored, including failures.

The formal interior flux consistency and the measured mesh stabilization
do not alone prove a uniform error bound for the continuum first
crossing. Such a proof would additionally control the evolution and
entropy errors, boundary treatment, and the transversality of the
continuum crossing. The present interval claims are therefore made for
specified finite grids only.

\subsection{Independent propagation and reproducibility}

For the selected eight-state network, modal and dense Pad\'e
matrix-exponential propagation agree on entropy-difference signs at
$t=1.26,1.28,1.56,2.54$; the largest absolute difference is
$4.42\times10^{-18}$. For the $N=64$ diffusion grid, the corresponding
checks at $t=3.08,3.11,4.03$ differ by at most $5.88\times10^{-16}$.
These checks use two different floating-point propagation methods.
Appendix~\ref{app:certification} gives the distinct interval arithmetic
calculation, its exact inputs, and the assertions it proves.

The supplied scripts use Python~3.12, NumPy~2.3.5, SciPy~1.17.0, and
Matplotlib~3.10.8. The interval program separately requires
\texttt{python-flint}~0.9.0. The main numerical driver regenerates all
draw-level tables, search results, mesh results, selected-model checks,
and the two-state negative control. The plotting driver reads these
saved results and produces the two vector figures. No unsuccessful
realization is deleted from the supplied data.

\section{Interval certification of the model statements}
\label{app:certification}

This appendix specifies the exact models, the interval algorithms, and
the distinction between a verified first-crossing enclosure and a
floating-point estimate. All the calculations described here are
self-contained in \texttt{certify\_generalization.py}; they do not require
data or code from an earlier manuscript. Table~\ref{tab:certified_scope}
summarizes the certified statements. No global uniqueness assertion is
made for these examples.

\begin{table*}[t]
\centering\small
\caption{Scope of the interval certificates. Temperatures are in units
with $T_b=1$. An interval for $t_M$ refers to the first meeting of relative
entropies, not a global uniqueness claim. The diffusion row concerns the
specified finite discretization.}
\label{tab:certified_scope}
\begin{tabular}{llll}
\toprule
Model & Preparation & First-meeting interval & Additional certificate\\
\midrule
Three states & $T_H=3,\ T_C=1.5$ & $(0.17022,0.17024)$ & $w_H>0$, $\Gamma>0$, $T_{\rm up}$\\
Three states & $T_H\in[2.9,3.1],\ T_C\in[1.45,1.55]$ & $(0.161,0.179)$ & Whole parameter rectangle\\
Eight-state network & $T_H=3,\ T_C=1.5$ & $(1.26,1.28)$ & Isolated spectrum, $\Gamma>0$, $T_{\rm up}$\\
Reflecting diffusion, $N=64$ & $T_H=3,\ T_C=1.5$ & $(3,3.2)$ & Finite-grid spectrum, $\Gamma>0$, $T_{\rm up}$\\
\bottomrule
\end{tabular}
\end{table*}

\subsection{Arithmetic and acceptance rules}
\label{app:cert_arithmetic}

The certificate uses Python~3.12.14, \texttt{python-flint}~0.9.0, and
FLINT~3.6.0. Arb represents a real quantity by a midpoint and an outward
error radius. The main run uses 384-bit midpoint precision; a separate
320-bit run passes the same proof obligations. Increasing precision is a
consistency check. The enclosure property of the arithmetic, together
with the analytic bounds below, is the certification argument.
The documented scalar and matrix interfaces provide interval elementary
functions and matrix exponentials.\footnote{The software interfaces are
documented at
\url{https://python-flint.readthedocs.io/en/latest/arb.html} and
\url{https://python-flint.readthedocs.io/en/latest/arb_mat.html}.}

Every physical input is supplied as an integer, exact fraction, or finite
decimal string. Decimal strings are first converted to exact rational
numbers. No binary floating-point number is used as a model parameter in
the certificate. Functions such as the potential and the jump rates are
then evaluated in ball arithmetic. A sign condition is accepted only if
the complete resulting interval is strictly on the required side of
zero. A check whose sign is unresolved stops the calculation. No small
eigenvalue, coefficient, probability, or imaginary part is replaced by
zero to obtain a successful test.

For an interval $X=[x^-,x^+]$, the scalar function
\begin{equation}
 g(x)=(1+x)\ln(1+x)-x
\end{equation}
decreases on $(-1,0)$ and increases on $(0,\infty)$, because
$g'(x)=\ln(1+x)$. Its interval range is therefore enclosed by endpoint
evaluation, including its exact minimum $g(0)=0$ whenever $0\in X$.
Similarly, $F(x)=g(x)/x^2$, with $F(0)=1/2$, is decreasing. Its enclosure
is obtained by evaluating the endpoints in reverse order. These
monotonicity rules reduce cancellation and interval overestimation; they
do not truncate a series or discard a remainder.

If intervals $X_i=[x_i^-,x_i^+]$ enclose several components, then
\begin{align}
\min_i x_i&\in[\min_i x_i^-,\min_i x_i^+],\\
\max_i x_i&\in[\max_i x_i^-,\max_i x_i^+].
\end{align}
The code uses these enclosure formulas, so it does not assume that the
index of an extremizing component is constant over a temperature cell.
For output, Arb's integer midpoint--radius decimal representation is
converted to explicit outward decimal endpoints. All shorter numerical
inequalities printed below are widened from those endpoints.

\subsection{Exact three-state model and its two-dimensional spectrum}
\label{app:cert_three_model}

The three states are labeled $0,1,2$, with exact energies and symmetric
kinetic prefactors
\begin{equation}
\begin{gathered}
 E=(0,2/5,3),\\
 (\gamma_{01},\gamma_{02},\gamma_{12})=(1,23/5,3/25).
\end{gathered}
\label{eq:cert_three_inputs}
\end{equation}
At the fixed final bath temperature $T_b=1$, define
\begin{equation}
\begin{gathered}
 k_{ij}=\gamma_{ij}e^{(E_i-E_j)/2}\quad(i\ne j),\\
 \pi_i=\frac{e^{-E_i}}{\sum_j e^{-E_j}}.
\end{gathered}
\end{equation}
The backward generator is $L_{ij}=k_{ij}$ for $i\ne j$ and
$L_{ii}=-\sum_{j\ne i}k_{ij}$. Probability columns evolve with $W=L^T$.
Every preparation at temperature $T$ is
\begin{equation}
 p_i(T)=\frac{e^{-E_i/T}}{\sum_j e^{-E_j/T}}.
\end{equation}
The final rates are independent of the preparation temperature.

The two nonstationary modes can be constructed analytically, without a
floating-point eigensolver. Set $q_i=\sqrt{\pi_i}$ and
\begin{equation}
 S=-\operatorname{diag}(q)L\operatorname{diag}(q)^{-1}.
\end{equation}
Detailed balance makes the exact matrix $S$ real symmetric, and $Sq=0$.
With $d_0=\sqrt{\pi_0+\pi_1}$, the vectors
\begin{equation}
\begin{gathered}
 u=\frac{(q_1,-q_0,0)}{d_0},\\
 v=\left(\frac{q_0q_2}{d_0},\frac{q_1q_2}{d_0},-d_0\right)
\end{gathered}
\end{equation}
are orthonormal and perpendicular to $q$. Define
$\alpha=u^TSu$, $\beta=u^TSv$, and $\delta=v^TSv$. The two positive
relaxation rates are exactly
\begin{equation}
 \lambda_{s,f}=\frac{\alpha+\delta
 \mp\sqrt{(\alpha-\delta)^2+4\beta^2}}{2}.
\label{eq:cert_two_rates}
\end{equation}
The certificate verifies $0<\lambda_s<\lambda_f$ and $\beta\ne0$.
A normalized backward eigenfunction for either rate is
\begin{equation}
 \phi_{n,i}=
 \frac{\beta u_i+(\lambda_n-\alpha)v_i}
 {q_i\sqrt{\beta^2+(\lambda_n-\alpha)^2}},\qquad n=s,f.
\label{eq:cert_two_modes}
\end{equation}
Thus $\sum_i\pi_i\phi_{n,i}=0$,
$\sum_i\pi_i\phi_{n,i}^2=1$, and
$L\phi_n=-\lambda_n\phi_n$. These identities follow from the
two-dimensional construction. Residual intervals containing zero are
additional implementation checks, rather than substitutes for that
construction.

For each preparation, calculate
\begin{equation}
\begin{aligned}
 b_n(T)&=\sum_i p_i(T)\phi_{n,i},\\
 f_i(t;T)&=\sum_{n=s,f}b_n(T)\phi_{n,i}e^{-\lambda_nt}.
\end{aligned}
\label{eq:cert_exact_deviation}
\end{equation}
The relative entropy is evaluated as
$D(t;T)=\sum_i\pi_i g(f_i(t;T))$. This expression is exact and has no
linear contribution that must cancel numerically.

\subsection{The fixed detuned preparation and arbitrary preparation errors}
\label{app:cert_detuned}

For $T_H=3$ and $T_C=3/2$, direct interval evaluation gives
\begin{equation}
\begin{gathered}
 1.81217<10^5w_H<1.81218,\\
 2.1699246260254279<10^4\Gamma,\\
 10^4\Gamma<2.1699246260254281,\\
 0.18492278483565589<T_{\rm up},\\
 T_{\rm up}<0.18492278483565590.
\end{gathered}
\end{equation}
Here $w_H=b_s(T_H)^2$, $w_C=b_s(T_C)^2$,
$q_H=b_f(T_H)^2$, and
$\Gamma=F(M_C(0))w_C-F(m_H(0))w_H$. The hot slow-mode weight is
strictly nonzero.

Write $d(t)=D_H(t)-D_C(t)$. Equation~\eqref{eq:cert_exact_deviation}
is evaluated over 1000 adjacent, closed, rational time intervals covering
$[0,0.169]$. Every interval gives $d>0$; the smallest certified lower
margin exceeds $7.4054\times10^{-6}$. To refine the crossing location,
use
\begin{equation}
\begin{aligned}
 \frac{dD}{dt}&=\sum_i\pi_i\ln(1+f_i)\dot f_i,\\
 \dot f_i&=-\sum_{n=s,f}\lambda_nb_n\phi_{n,i}e^{-\lambda_nt}.
\end{aligned}
\end{equation}
Thirty adjacent intervals cover $[0.169,0.172]$ and verify $d'<0$ on
every cell. A separate evaluation of $e^{Wt}p^H(0)$ and $e^{Wt}p^C(0)$
then gives
\begin{equation}
\begin{aligned}
 d(0.17022)&>3.9872\times10^{-8},\\
 d(0.17024)&<-8.1501\times10^{-8}.
\end{aligned}
\end{equation}
The early positivity excludes any earlier meeting, and the local
derivative check excludes an earlier root within the final bracket.
Consequently $0.17022<t_M<0.17024$. The derivative claim is local;
it is not extended to all later times.

The same exact nominal preparations permit a conservative error
certificate. Let $\widetilde p^\alpha(0)$ be any normalized probability
vector satisfying $\|\widetilde p^\alpha(0)-p^\alpha(0)\|_1\le10^{-6}$,
independently for $\alpha=H,C$. The code verifies componentwise that
\begin{equation}
\begin{gathered}
 0.7<\frac{p_i^\alpha(0)-10^{-6}}{\pi_i},\\
 \frac{p_i^\alpha(0)+10^{-6}}{\pi_i}<6.
\end{gathered}
\end{equation}
The original and perturbed likelihood ratios therefore remain in
$[0.7,6]$ at all later times by the Markov maximum principle. Centering
the entropy gradient on this range gives the Lipschitz constant
$C=\frac12\ln(6/0.7)<1.074218$. The combined error in $d$ is below
$2C\times10^{-6}<2.148436\times10^{-6}$. This is less than the verified
positive margin on $[0,0.169]$ and the magnitude of
$d(0.172)<-1.03253\times10^{-5}$. Every permitted pair consequently has
its first meeting in $(0.169,0.172)$. The tolerance is sufficient; no
maximal tolerance is claimed.

\subsection{Certification of an entire temperature rectangle}
\label{app:cert_temperature_rectangle}

Now allow both preparations to vary throughout the closed rectangle
\begin{equation}
 T_H\in[2.9,3.1],\qquad T_C\in[1.45,1.55],\qquad T_b=1.
\label{eq:cert_temperature_box}
\end{equation}
This is a continuum of initial conditions. The calculation covers the
hot interval with the 200 closed cells
\begin{equation}
 I_j^H=\left[\frac{29}{10}+\frac{j}{1000},
                    \frac{29}{10}+\frac{j+1}{1000}\right],
 \quad j=0,\ldots,199,
\end{equation}
and the cold interval with the 100 closed cells
\begin{equation}
 I_j^C=\left[\frac{29}{20}+\frac{j}{1000},
                    \frac{29}{20}+\frac{j+1}{1000}\right],
 \quad j=0,\ldots,99.
\end{equation}
All endpoints are exact rationals. The first and last endpoints, the
common width, and equality of every pair of adjacent endpoints are
asserted explicitly. Thus there are no gaps, including at the boundary
of the rectangle.

For each temperature cell, interval evaluation of the Gibbs formula and
Eq.~\eqref{eq:cert_exact_deviation} encloses every probability and modal
coefficient throughout that cell. Taking an interval hull over all hot
cells and, separately, all cold cells gives bounds valid for every pair
in the product rectangle. This independent enclosure of the two
preparations can be conservative, but it cannot omit a pair. In
particular, the certificate gives uniformly
\begin{equation}
\begin{gathered}
 w_H>3.24019\times10^{-6},\\
 \Gamma>1.8624069\times10^{-4},\\
 T_{\rm up}<0.190298.
\end{gathered}
\label{eq:cert_uniform_margin}
\end{equation}
No hot preparation in this rectangle has an exactly canceled slow mode.

The all-mode envelopes improve the time interval. For compactness in this
subsection only, write $H=b_s(T_H)^2$, $A=b_f(T_H)^2$,
$C=b_s(T_C)^2$, $B=b_f(T_C)^2$, and $\Delta=\lambda_f-\lambda_s$.
At a certified reference $s$, let
$a_\alpha=F(M_\alpha(s))$ and $c_\alpha=F(m_\alpha(s))$.
For every $t\ge s$ the exact lower comparison is
\begin{equation}
 e^{2\lambda_st}d(t)\ge
 (a_HH-c_CC)+(a_HA-c_CB)e^{-2\Delta t}.
\label{eq:cert_family_lower}
\end{equation}
The coefficient $a_HA-c_CB$ is strictly positive throughout the whole
temperature rectangle at each reference used below. For each fixed pair
of temperatures, the right-hand side of
Eq.~\eqref{eq:cert_family_lower} is therefore decreasing with $t$.
If its interval enclosure is positive at $t=L$, positivity is proved for
every $t\in[s,L]$, not merely at $L$. Table~\ref{tab:temperature_certificate}
gives the four successive reference intervals. Induction covers all
times from zero through $0.161$.

\begin{table}[t]
\centering
\caption{Uniform temperature-rectangle certificate. The last two columns
are conservative lower bounds on the fast coefficient in
Eq.~\eqref{eq:cert_family_lower} and on its right-hand side at $t=L$.
Every entry applies to the entire rectangle
\eqref{eq:cert_temperature_box}.}
\label{tab:temperature_certificate}
\begin{tabular}{rrrr}
\toprule
$s$ & $L$ & $a_HA-c_CB>$ & Scaled lower margin $>$\\
\midrule
0 & 0.140 & 0.10319 & $3.61936\times10^{-5}$\\
0.140 & 0.155 & 0.24883 & $1.67882\times10^{-4}$\\
0.155 & 0.160 & 0.25419 & $8.64680\times10^{-5}$\\
0.160 & 0.161 & 0.25567 & $7.25007\times10^{-5}$\\
\bottomrule
\end{tabular}
\end{table}

The upper comparison at the last reference is
\begin{equation}
 e^{2\lambda_st}d(t)\le
 (c_HH-a_CC)+(c_HA-a_CB)e^{-2\Delta t}.
\end{equation}
Its interval upper endpoint at $t=0.179$ is below
$-3.12237\times10^{-5}$. Hence $d(0.179)<0$ for every pair in the
rectangle. Combining this sign with the absence of an earlier meeting
proves
\begin{equation}
 \boxed{0.161<t_M(T_H,T_C)<0.179}
\label{eq:cert_family_crossing}
\end{equation}
throughout the closed rectangle. In particular, its interior provides
an open set of preparations with a certified crossing.

\subsection{A general certificate excluding earlier crossings}
\label{app:cert_time_cells}

For the larger examples we avoid a spectral assumption in the
first-crossing computation. The following bound turns interval
propagation at cell endpoints into a statement about every intervening
time. Let $P(u)=e^{Wu}$ and let $z$ be any signed vector. Positivity and
unit column sums imply
\begin{equation}
 \|P(u)z\|_1
 \le\sum_{i,j}P_{ij}(u)|z_j|
 =\sum_j|z_j|=\|z\|_1.
\end{equation}
At a left endpoint $t_j$, define
$v_\alpha=\|Wp^\alpha(t_j)\|_1$. Because $W$ commutes with its
semigroup,
\begin{equation}
 \|\dot p^\alpha(t_j+u)\|_1
 =\|P(u)Wp^\alpha(t_j)\|_1\le v_\alpha.
\end{equation}
Integration for $0\le u\le\delta t$ gives
$\|p^\alpha(t_j+u)-p^\alpha(t_j)\|_1\le u v_\alpha$.
The difference has zero total mass, so the absolute displacement of
one component is at most half its $L^1$ norm. Consequently
\begin{equation}
 \frac{p_i^\alpha(t_j)-\delta t\,v_\alpha/2}{\pi_i}
 \le\frac{p_i^\alpha(t_j+u)}{\pi_i}
 \le\frac{p_i^\alpha(t_j)+\delta t\,v_\alpha/2}{\pi_i}.
\label{eq:cert_cell_ratios}
\end{equation}
The minimum lower endpoint and maximum upper endpoint in
Eq.~\eqref{eq:cert_cell_ratios} define $h_{\alpha,-}$ and
$h_{\alpha,+}$. The code requires $h_{\alpha,-}>0$.

To bound the entropy change, join the two probability vectors by a
straight segment. Its likelihood ratios stay in this same range. The
gradient of $D$ has components $\ln(p_i/\pi_i)+1$. Subtracting the
midpoint of the gradient range does not change its contraction with a
zero-mass probability difference. Therefore, with
\begin{equation}
 C_\alpha=\frac12\ln\frac{h_{\alpha,+}}{h_{\alpha,-}},
\end{equation}
integration along that segment gives
\begin{equation}
\begin{aligned}
 &|D_\alpha(t_j+u)-D_\alpha(t_j)|\\
 &\quad\le C_\alpha\|p^\alpha(t_j+u)-p^\alpha(t_j)\|_1\\
 &\quad\le\delta t\,C_\alpha v_\alpha.
\end{aligned}
\end{equation}
Thus the computable cell margin
\begin{equation}
 \mathcal M_j=d(t_j)
 -\delta t(C_Hv_H+C_Cv_C)
\label{eq:cert_cell_margin}
\end{equation}
is a lower bound for $d(t)$ throughout $[t_j,t_j+\delta t]$.
All quantities on the right are interval evaluated; a strictly positive
lower endpoint certifies the entire cell.

For each uniform partition the program computes the interval matrix
$P(\delta t)$ once, then propagates both interval probability columns
by repeated multiplication. Dependency between components may widen an
enclosure but does not remove the exact trajectory. Exact rational
partition endpoints establish complete time coverage. The exact target
probability vectors remain normalized by the generator construction;
the proof does not assume that the component sum of a finite-width
interval vector equals the point interval $[1,1]$.

\begin{table}[t]
\centering
\caption{Absence of an earlier crossing for the larger fixed models.
The margin is a lower bound for every time in each cell, including the
cell boundaries. A negative sign at the separate right endpoint then
gives the stated first-crossing interval.}
\label{tab:cert_time_cells}
\begin{tabular}{lrrr}
\toprule
Model & Positive interval & Cells & $\min_j\mathcal M_j>$\\
\midrule
Network, $N=8$ & $[0,1.26]$ & 1000 & $1.5774\times10^{-6}$\\
Diffusion, $N=64$ & $[0,3]$ & 3000 & $1.33238\times10^{-5}$\\
\bottomrule
\end{tabular}
\end{table}

\subsection{Exact larger-model inputs and certified spectral margins}
\label{app:cert_larger_models}

For the selected eight-state network, the exact energies are
\begin{equation}
 E=\frac{1}{7}(25,10,15,0,35,30,5,20).
\label{eq:cert_eight_energies}
\end{equation}
Table~\ref{tab:cert_eight_prefactors} supplies all 28 independent
prefactors; $\gamma_{ji}=\gamma_{ij}$ and $\gamma_{ii}=0$.
The printed finite decimals define exact rational prefactors in the
certificate. They are the recorded prefactors of the selected numerical
sample. The certificate uses the exact rational energy grid rather than
the binary approximations used in exploratory diagonalization. These
definitions fix the model without requiring access to a random-number
generator or its internal state. Rates are again
$k_{ij}=\gamma_{ij}\exp[(E_i-E_j)/2]$, with $T_b=1$ and preparations
at $T_H=3$, $T_C=3/2$.

\begin{table*}[t]
\centering\small
\caption{Exact finite-decimal prefactors of the certified eight-state
network. State labels run from 0 to 7; reverse-edge prefactors are equal.}
\label{tab:cert_eight_prefactors}
\begin{tabular}{cr@{\qquad}cr}
\toprule
Edge $(i,j)$ & $\gamma_{ij}$ & Edge $(i,j)$ & $\gamma_{ij}$\\
\midrule
(0,1) & 0.9188449204806782 & (2,4) & 0.023278092516836695 \\
(0,2) & 1.8270639941702176 & (2,5) & 0.179923223395374 \\
(0,3) & 3.0757387290827336 & (2,6) & 0.09150247227787736 \\
(0,4) & 0.3983756194312874 & (2,7) & 0.21305578086689048 \\
(0,5) & 0.554196771727935 & (3,4) & 0.3119478763395427 \\
(0,6) & 0.030975444662106826 & (3,5) & 1.695725324502788 \\
(0,7) & 5.504807369055709 & (3,6) & 0.025315305950964238 \\
(1,2) & 0.26547539647973406 & (3,7) & 4.302106617048917 \\
(1,3) & 0.13466630019339842 & (4,5) & 0.1381049472097854 \\
(1,4) & 3.247924074852203 & (4,6) & 6.536342154044589 \\
(1,5) & 3.9288672894123398 & (4,7) & 0.04819180074308499 \\
(1,6) & 0.09045980833757165 & (5,6) & 0.19478346839791302 \\
(1,7) & 0.3404994855323166 & (5,7) & 4.292828540168284 \\
(2,3) & 4.835837000620058 & (6,7) & 0.0647499787650152 \\
\bottomrule
\end{tabular}
\end{table*}

For the diffusion certificate, take $N=64$, $\delta x=6/N$, and
$x_i=-3+(i+1/2)\delta x$, $i=0,\ldots,N-1$. Define the potential
exactly by
\begin{equation}
 U(x)=-3\exp\left[-\frac{(x+1)^2}{2(1/5)^2}\right]
      -2\exp\left[-\frac{(x-1)^2}{2(7/10)^2}\right].
\label{eq:cert_diffusion_potential}
\end{equation}
Only adjacent cells communicate, with
\begin{equation}
 k_{i,i+1}=\delta x^{-2}
   \exp\left[-\frac{U(x_{i+1})-U(x_i)}{2}\right],
\end{equation}
and the reverse rate obtained by exchanging the indices. No jump leaves
the domain. Every coordinate, width, and potential parameter is rational
before the exponentials are evaluated with Arb. The stationary
probabilities are proportional to $e^{-U(x_i)}$; the preparation weights
are proportional to $e^{-U(x_i)/T}$ at $T=3,3/2$.

For both larger fixed models the general theorem's inputs are also
certified. Form the real symmetric matrix
$S=-\operatorname{diag}(\sqrt\pi)L\operatorname{diag}(\sqrt\pi)^{-1}$.
The program calls the validated Rump eigensolver with eigenvectors
requested. It requires a successful isolation of all eigenvalues and
checks that the real-part intervals are strictly ordered, with an
interval containing zero followed by strictly positive intervals. The approximate-only
eigensolver option is never used. The validated routine encloses actual
eigenpairs; a small residual alone would not supply this guarantee.
The corresponding software specification is given in the documented
\texttt{acb\_mat.eig} interface.\footnote{
\url{https://python-flint.readthedocs.io/en/latest/acb_mat.html};
the validated eigenpair algorithms are described at
\url{https://arblib.org/acb_mat.html}.}

The exact matrix is real symmetric, so its eigenvalues are real. The real
part of an enclosed complex eigenvector is a real eigenvector for the
same exact real eigenvalue. Its Euclidean norm is certified strictly
positive before division. Normalizing that vector and dividing its
components by $\sqrt{\pi_i}$ gives the weighted-normalized slow
eigenfunction. The spectral weights $w_H,w_C$ then follow from the Gibbs
projections. The remaining hot weight is computed exactly as
\begin{equation}
 q_H=\sum_i\pi_i\left(\frac{p_i^H(0)}{\pi_i}-1\right)^2-w_H.
\end{equation}
The isolated spectrum verifies that the slow eigenspace is
one-dimensional in these two particular models. The general theorem
itself allows degeneracy when the whole slow subspace is used.

\begin{table*}[t]
\centering
\caption{Conservative outward intervals for the spectral sufficient
condition in the two larger fixed models. The full eigenvalue and
eigenvector-derived enclosures are in the machine-readable certificate.}
\label{tab:cert_larger_spectrum}
\begin{tabular}{lll}
\toprule
Quantity & Network, $N=8$ & Diffusion, $N=64$\\
\midrule
$w_H$ & $(4.14667,4.14668)10^{-6}$ & $(1.10687,1.10688)10^{-5}$\\
$\Gamma$ & $(1.2529348,1.2529350)10^{-4}$ & $(4.0356790,4.0356791)10^{-4}$\\
$T_{\rm up}$ & $(2.53950003286111,\,2.53950003286112)$ & $(4.02143847094726,\,4.02143847094728)$\\
\bottomrule
\end{tabular}
\end{table*}

Direct, separately computed matrix exponentials give
\begin{align}
 d_8(1.26)&>2.0557824\times10^{-6},\\
 d_8(1.28)&<-2.1815267\times10^{-6},\\
 d_{64}(3)&>1.3766163\times10^{-5},\\
 d_{64}(3.2)&<-1.2647043\times10^{-5}.
\end{align}
Combined with Table~\ref{tab:cert_time_cells}, these signs establish
$1.26<t_M^{(8)}<1.28$ and $3<t_M^{(64)}<3.2$. The first-crossing
calculation therefore has a propagation-based verification independent
of the spectral-margin calculation.

\subsection{A certified upper endpoint using two retained modes}
\label{app:cert_partial}

The selected eight-state model also permits a direct certificate for the
partial-spectrum construction. In the conventional ordering
$0=\lambda_1<\lambda_2<\lambda_3<\lambda_4<\cdots$, the validated
eigensolver gives
\begin{equation}
 17.27305635<\lambda_4<17.27305637.
\end{equation}
Choose the exact rational lower cutoff
$\underline\lambda_4=17.27305635$, and retain only the two coefficients
$b_{2,\alpha}$ and $b_{3,\alpha}$ in each preparation. Their certified
eigenfunctions are normalized as above. Define the remaining hot norm
\begin{equation}
 R_H(0)=\chi^2(p_H(0)\Vert\pi)-b_{2,H}^2-b_{3,H}^2>0.
\end{equation}
Orthogonality and the cutoff imply
\begin{equation}
\begin{aligned}
 Q_H(t)&\le\overline Q_H(t),\\
 \overline Q_H(t)&=b_{2,H}^2e^{-2\lambda_2t}
                 +b_{3,H}^2e^{-2\lambda_3t}\\
 &\quad+R_H(0)e^{-2\underline\lambda_4t}.
\end{aligned}
\end{equation}
\begin{equation}
\begin{aligned}
 Q_C(t)&\ge\underline Q_C(t),\\
 \underline Q_C(t)&=b_{2,C}^2e^{-2\lambda_2t}
                   +b_{3,C}^2e^{-2\lambda_3t}.
\end{aligned}
\end{equation}
In particular, the omitted hot contribution is bounded, rather than
set to zero. At the exact time $t=155163/100000=1.55163$, the interval
calculation gives
\begin{equation}
\begin{gathered}
 c_H\overline Q_H(t)-a_C\underline Q_C(t)\\
 {}\in(-8.000113,-8.000112)\times10^{-10}.
\end{gathered}
\end{equation}
The right-hand side is an upper bound on $d(t)$, so continuity and the
verified initial ordering imply the certified upper endpoint
$t_M<1.55163$. This sign test certifies an endpoint, not the numerical
root of the envelope or uniqueness of the entropy crossing. It also
does not establish a computational speedup: the full eigensystem was
computed to validate the cutoff and retained projections in this
demonstration.

\subsection{Reproduction and limits of the certificates}
\label{app:cert_reproduction}

Install the pinned dependency in \texttt{validation/requirements.txt},
then execute \texttt{certify\_generalization.py}. The options
\texttt{--bits 384} and \texttt{--bits 320} reproduce the two recorded
runs. Python's optimization flag \texttt{-O} must not be used because
the assertion statements are explicit proof obligations. The output
records the software versions, source and exact-input hashes, spectral
intervals, parameter-cell counts, reference-stage coefficients, time-cell
counts, smallest positive margins, and endpoint signs. An unresolved
interval or failed assumption terminates the program.

The continuous temperature-rectangle result and the stated fixed-model
inequalities are certified. The finite network-ensemble classifications,
the numerical error and coverage summaries across that ensemble, and the
other diffusion mesh sizes are ordinary floating-point results unless
explicitly identified otherwise. In particular, successful mesh
refinement is not a rigorous continuum-error bound. Neither the network
sampling frequencies nor these selected model certificates imply that
Mpemba crossings are common under every sampling law, or that the
analytic bounds are uniformly sharp. The algorithms combine standard
entropy comparisons, contraction, validated eigenpairs, and interval
propagation; no priority claim for those individual ingredients is made.

\end{document}